\documentclass[useAMS,usenatbib,usegraphicx]{mn2e}
\usepackage{amsmath}
\usepackage{color}

\newif\ifAMStwofonts
\AMStwofontstrue

\pdfoutput=1
\newcommand{\simlt}{\lower.5ex\hbox{$\; \buildrel < \over \sim \;$}}
\newcommand{\simgt}{\lower.5ex\hbox{$\; \buildrel > \over \sim \;$}}
\newcommand{\kms}{km\,s$^{-1}$}

\title[GAMA: Merger histories via stellar populations]
{Galaxy and Mass Assembly (GAMA):
Probing the merger histories of massive galaxies via stellar populations}
\author[I. Ferreras et al.]
{I. Ferreras$^1$\thanks{E-mail: i.ferreras@ucl.ac.uk}, 
A.~M. Hopkins$^2$, M.~L.~P. Gunawardhana$^{3,4}$, A.~E. Sansom$^5$, \and
M.~S. Owers$^{2,6}$, S. Driver$^{7,8}$, L. Davies$^7$, A. Robotham$^7$, \and
E.~N. Taylor$^9$, I. Konstantopoulos$^2$, S. Brough$^2$, P. Norberg$^3$, \and
S. Croom$^{10,11}$, J. Loveday$^{12}$, L.~Wang$^{13,14}$, M. Bremer$^{15}$\\
\\
$^1$ Mullard Space Science Laboratory, University College London, 
Holmbury St Mary, Dorking, Surrey RH5 6NT, UK\\
$^2$ Australian Astronomical Observatory, PO Box 915, North Ryde, NSW 1670, Australia\\
$^3$ ICC \& CEA, Department of Physics,
Durham University, South Road, Durham DH1 3LE, UK\\
$^4$ Pontificia Universidad Cat\'olica de Chile, Vicu\~na Mackenna 4860, 7820436 Macul, Santiago, Chile\\
$^5$ Jeremiah Horrocks Institute, University of Central Lancashire, Preston PR1 2HE, UK\\
$^6$ Department of Physics and Astronomy, Macquarie University, Sydney, NSW 2109, Australia\\
$^7$ International Centre for Radio Astronomy Research, The University of Western Australia,
35 Stirling Hwy, Crawley, WA 6009, Australia\\
$^8$ SUPA, School of Physics \& Astronomy, University of St Andrews, North Haugh,
St Andrews KY16 9SS, UK\\
$^9$ School of Physics, David Caro Building, The University of Melbourne, Parkville VIC 3010, Australia\\
$^{10}$ Sydney Institute for Astronomy (SIfA), School of Physics,
The University of Sydney, NSW 2006, Australia\\
$^{11}$ ARC Centre of Excellence for All-sky Astrophysics (CAASTRO),
44-70 Rosehill Street, Redfern NSW 2016, Sydney, Australia\\
$^{12}$ Astronomy Centre, University of Sussex, Falmer, Brighton BN1 9QH, UK\\
$^{13}$ SRON Netherlands Insitute for Space Research, Landleven 12, 9747 AD, Groningen, the Netherlands\\
$^{14}$ Kapteyn Astronomical Institute, University of Groningen, Postbus 800, 9700 AV Groningen, the Netherlands\\
$^{15}$ H H Wills Physics Laboratory, Tyndall Avenue, Bristol BS8 1TL, UK
}

\begin{document}
\date{MNRAS, accepted 2017 February 24}
\pagerange{\pageref{firstpage}--\pageref{lastpage}} \pubyear{2017}
\maketitle
\label{firstpage}


\begin{abstract}
The merging history of galaxies can be traced with studies of
dynamically close pairs.  These consist of a massive primary galaxy
and a less massive secondary (or satellite) galaxy. The study of the
stellar populations of secondary (lower mass) galaxies in close pairs
provides a way to understand galaxy growth by mergers. Here we focus
on systems involving at least one massive galaxy -- with stellar mass
above $10^{11}$M$_\odot$ in the highly complete GAMA survey.  Our working sample
comprises 2,692 satellite galaxy spectra ($0.1\leq z\leq 0.3$). These spectra
are combined into high S/N stacks, and binned according to both an
``internal'' parameter, the stellar mass of the satellite galaxy
(i.e. the secondary), and an ``external'' parameter, selecting either
the mass of the primary in the pair, or the mass of the corresponding
dark matter halo. We find significant variations in the age of the
populations with respect to environment. At fixed mass, satellites
around the most massive galaxies are older and possibly more metal
rich, with age differences $\sim$1--2\,Gyr within the subset of lower mass
satellites ($\sim 10^{10}$M$_\odot$). These variations are similar when stacking
with respect to the halo mass of the group where the pair is embedded.
The population trends in the lower-mass satellites are consistent with
the old stellar ages found in the outer regions of massive galaxies.
\end{abstract} 

\begin{keywords}
galaxies: evolution -- galaxies: formation -- galaxies: interactions -- galaxies: stellar content.
\end{keywords}

\section{Introduction}
\label{Sec:Intro}

Galaxy growth is one of the fundamental processes linking
structure formation and the observable Universe. The connection
between the evolution of (dark matter-dominated) structures and the
``baryon physics'' of galaxy formation is the ``holy grail'' of
extragalactic astrophysics.  Galaxy mergers lead both to 
the mixing of the stellar component already in place
in the progenitors, and to gas inflows that provide newly formed
stars. In particular, massive galaxies are one of the best targets to
put these processes to the test. At present, the most widely accepted theory
for the formation of massive galaxies relies on a two-step
scenario \citep[e.g.,][]{Oser:12}. An early stage of collapse and
efficient star formation {\sl in-situ} builds the core of massive galaxies,
whereas the outer regions are populated by stars formed {\it ex situ},
incorporated in the galaxy during subsequent merging events. This
theoretical scenario was motivated by the observations of compact,
massive galaxies at high redshift
\citep[see, e.g.,][]{Daddi:05,NT:06,NT:07,vD:10}, in contrast with
their more extended counterparts at lower redshift.  Studies of the
stellar populations in these compact systems reveal a growth mechanism
driven by the accretion of the stellar components of companion
galaxies through merging, ruling out significant levels of in situ star
formation during this growth phase \citep{I3}.

Simulations provide grounds for this
interpretation \citep{Hirschmann:15}, although there are still important
aspects, such as the radial gradients in the age and chemical
composition of the populations within galaxies, that require more
work. Whilst the outer regions of massive galaxies are expected to
originate mostly from minor mergers \citep{Naab:09}, the old stellar
ages found in the outskirts of massive early-type
galaxies \citep{FLB:12,Greene:15} reveal an important environment-related effect
on the progenitor (minor-merging) systems. The stellar populations of
satellites located dynamically close to massive galaxies are expected to be
incorporated into the merged system.  Therefore, a simple but
effective way to study merging systems is based on
samples of galaxies involving very close pairs in projected distance
and relative velocity, i.e. those that would be expected to merge
within a relatively short time \citep[see,
e.g][]{Patton:00,LeFevre:00,Ben:09,Newman:12,CLS:12,EMQ:12,EMQ:13,Conselice:14}.
In \citet{SH4} a sample of close pairs at redshift z$\sim$0.5-1 was
explored via medium-band photometry \citep[SHARDS,][]{SHARDS}, serving
as a low-resolution (R$\sim$50) spectrograph. The age of the
populations in galaxies dynamically close to massive galaxies was
found to obey the same mass-age relationship as galaxies in the
field. That study was complete down to a 1:30 stellar mass ratio,
suggesting that such a scenario would not be compatible with the flat
age gradients found in massive ETGs, unless the net mass fraction
provided by minor mergers was small. Furthermore, a number count
analysis supported the idea that mass ratios closer to $\sim$1:3
dominate this growth channel in massive galaxies. An extension of
this study to a very large sample of low-z galaxies from
SDSS \citep{Ruiz:14} confirmed that mass ratios in the region 1:5 are
more important than minor merging systems (typically defined by mass
ratios below 1:10).

This project focuses on a more accurate determination of the
properties of stellar populations, taking advantage of the high
density of optical spectra available in GAMA (with respect to SDSS) to
build a catalogue of galaxy spectra comprising dynamically close pairs
with at least one massive galaxy (the DR2 GAMA catalogue already
includes over 15,000 galaxies with stellar mass above
$10^{11}$M$_\odot$). These pairs are the progenitors of merged
systems, and via comparisons with numerical simulations \citep[see,
e.g.,][]{KW:08,Jiang:14} it is possible to derive the actual merger
rates as a function of the mass ratio.  Moreover, the selection of
pairs involving at least a massive galaxy probes the regime where most
of the growth proceeds via the accretion of lower mass
galaxies \citep{Robotham:14}.  This paper aims at measuring the
properties of the stellar populations of the merger progenitors with
respect to stellar mass, mass ratio, and environment.  The targeted
close pairs are expected to merge at later times, mixing up their
stellar populations.  Depending on the dynamical characteristics of
the system (mainly the merger mass ratio), it is possible to infer the
properties of the radial gradients of stellar population properties in
massive galaxies at later times \citep[e.g.,][]{FLB:11}. This work
complements the recent GAMA-based study
of \citet{Davies:Pairs,Davies:16}, devoted to the trends of star
formation diagnostics in close pairs.

The methodology involves a careful stacking of observed spectra --
binned according to common characteristics such as stellar mass or
merger mass ratio -- in order to reach the high SNR needed to explore
a set of line strength indices dependent on the age, metallicity and
[$\alpha$/Fe] of the populations. We note that due to flux calibration
issues in the GAMA/AAT spectra, it is better to probe the stellar
populations via indices, avoiding spectral fitting (except for the
narrow spectral window straddling the 4000\,\AA\ break).  Note that
the level of merging, derived from the analysis of close pairs, gives
a fairly constant rate out to z$\sim$1.5, quantified as a stellar mass
growth inverse timescale of $\tau^{-1}\equiv(\Delta M/M)/\Delta t=
0.08\pm 0.02$\,Gyr$^{-1}$
\citep{SH4}. Over the redshift probed in this sample ($z\leq 0.3$), one would
thus expect a fractional stellar mass growth from mergers between 10\% (at
z=0.1) and 30\% (at z=0.3). The analysis of the stellar populations of
the satellite galaxies will impose valuable constraints on models of
galaxy formation and evolution \citep[e.g.,][]{Hirschmann:15}.

\begin{figure}
\begin{center}
\includegraphics[width=8.4cm]{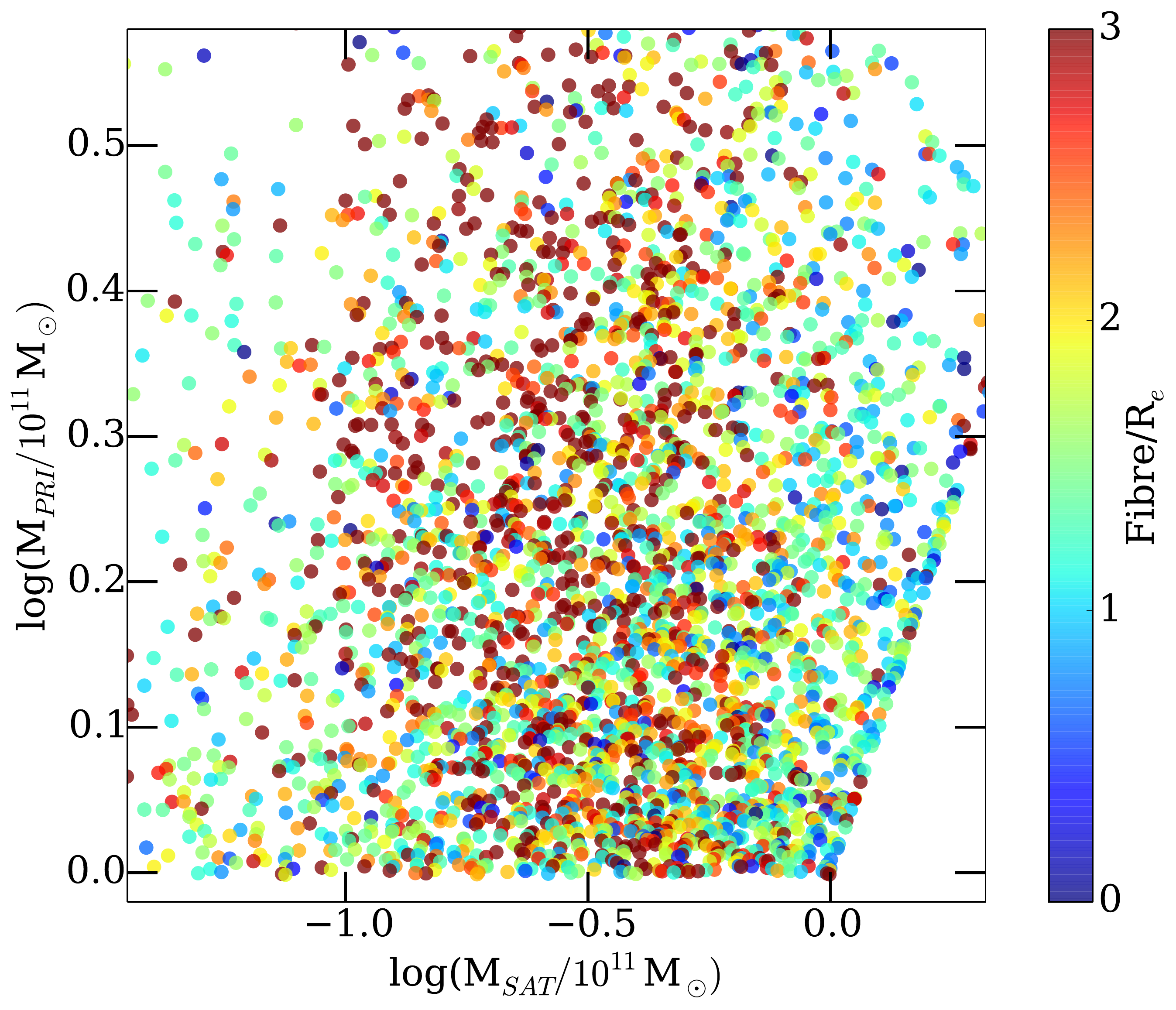}
\includegraphics[width=8.4cm]{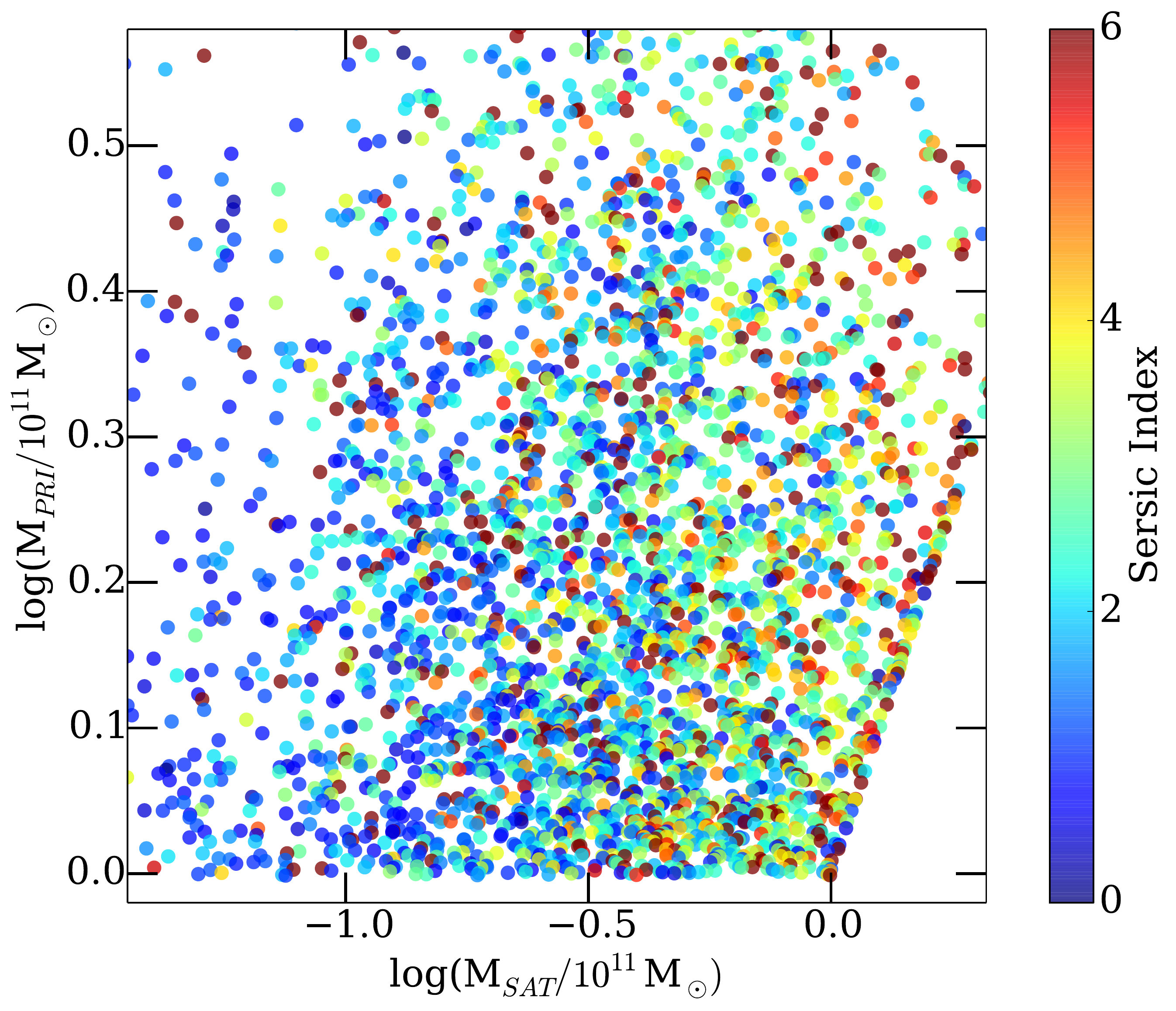}
\includegraphics[width=8.7cm]{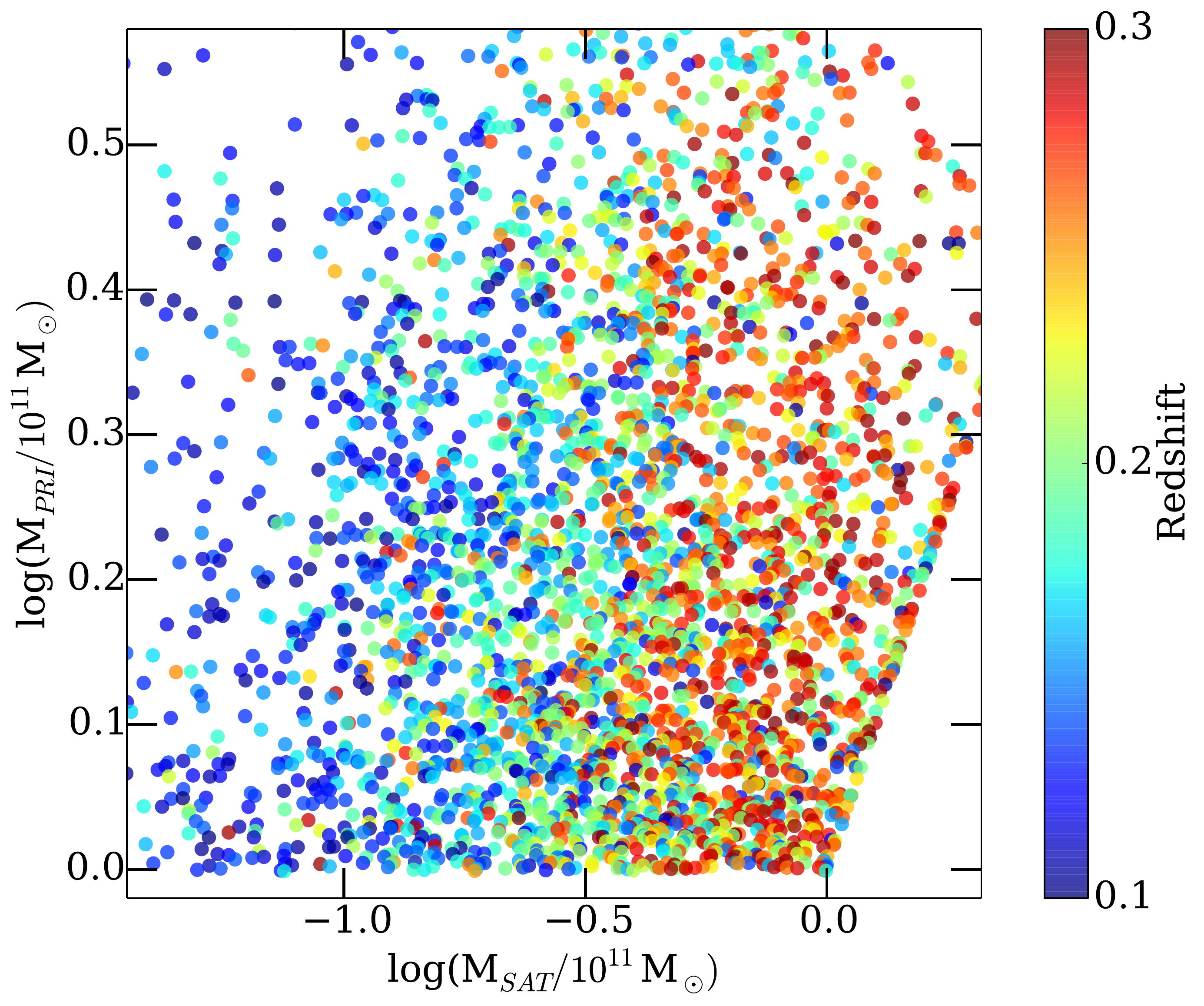}
\end{center}
\caption{Colour-coded plots representing the sample on the
primary vs secondary stellar mass. Colour represents fibre aperture
(in units of the effective radius, {\sl top});
S\'ersic index ({\sl middle}); and redshift ({\sl bottom}).
Note that a bias with respect to these quantities would show
up as a significant colour trend (see text for details).
}
\label{fig:RenSz}
\end{figure}

The structure of the paper is as follows: \S\ref{Sec:Data} presents
the data set extracted from the GAMA survey.
\S\ref{Sec:Prepare} describes how the spectra are prepared and stacked, followed
by \S\ref{Sec:Method}, devoted to the methodology regarding
the analysis of the line strengths. A discussion section
(\S\ref{Sec:Disc}) puts the results in context with our understanding
of galaxy growth mechanisms, concluding with a summary in \S\ref{Sec:Summ}.
A standard $\Lambda$CDM cosmology is
adopted, with $\Omega_m=0.27$ and
H$_0=70$\,km\,s$^{-1}$Mpc$^{-1}$. For reference, the look-back time to
z=0.2 (the median of our working sample) is 2.4\,Gyr and 1\,arcsec
maps into 3.3\,kpc at that redshift.

\begin{figure}
\begin{center}
\includegraphics[width=8.9cm]{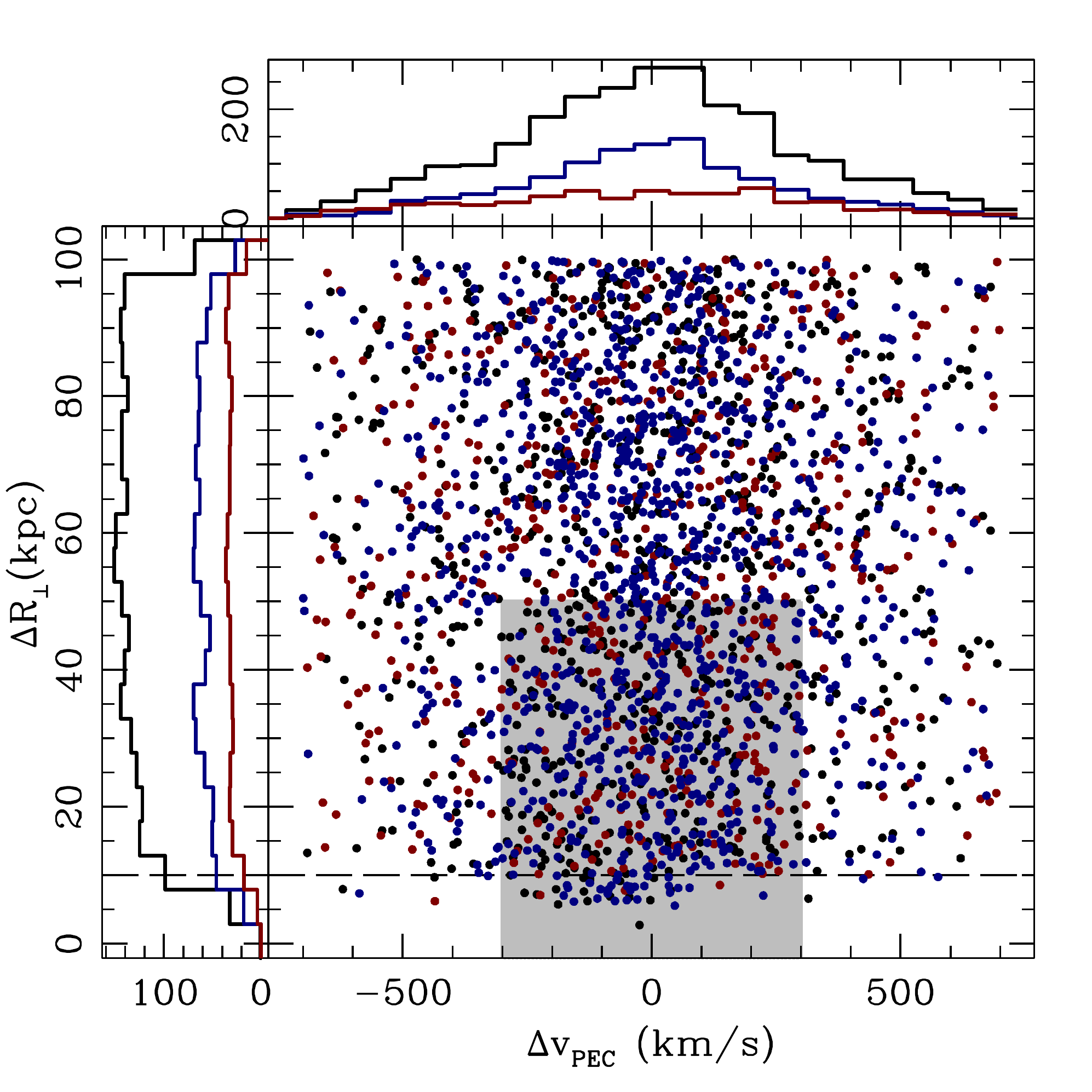}
\end{center}
\caption{Distribution of the sample with respect to
relative velocity ($\Delta$v$_{\rm PEC}$) and projected physical
separation ($\Delta$R$_\perp$).  The red (blue) dots and histograms
correspond to the subsample of satellites around the highest (lowest)
mass primary galaxies. The horizontal dashed line marks the
$\Delta$R=10\,kpc lower limit imposed on the satellite
sample.  The shaded region defines an additional subset of tighter
galaxy pairs (see text for details).}
\label{fig:vDplot}
\end{figure}

\begin{figure*}
\begin{center}
\includegraphics[width=170mm]{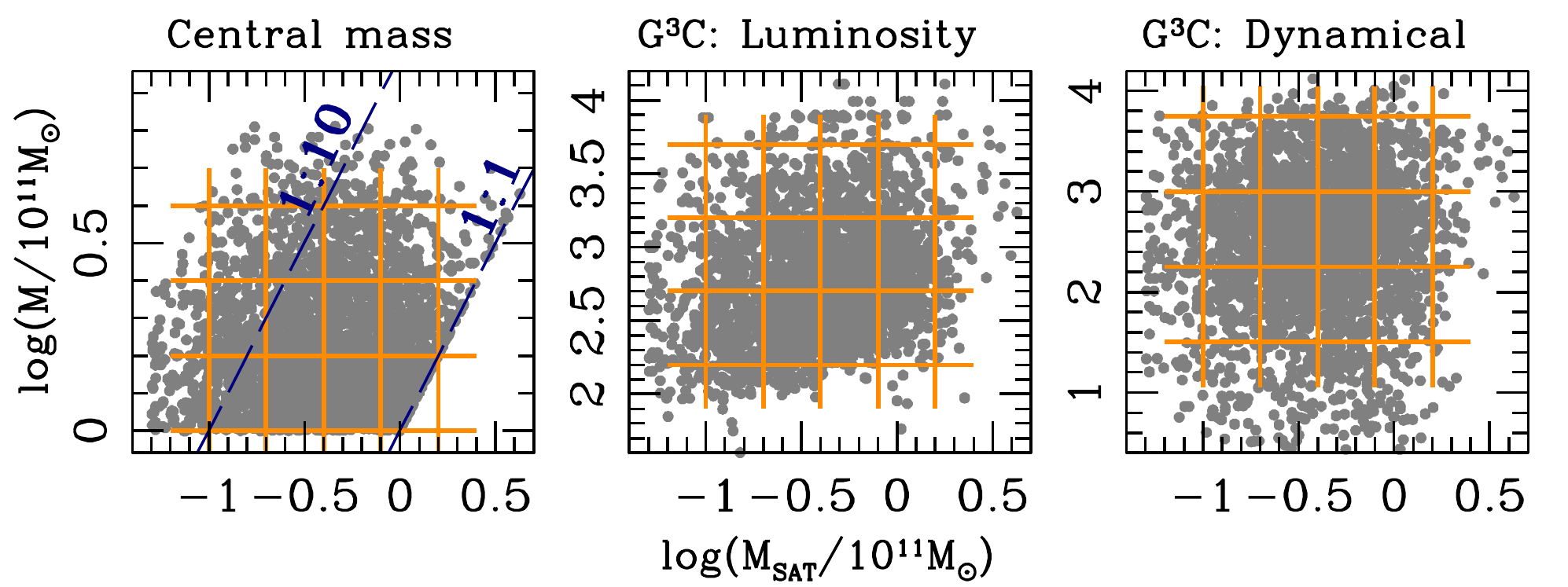}
\end{center}
\caption{Grids defining the stacking procedure: This diagram shows the stellar
mass of the primary and secondary galaxies for each of the close pairs
defined in the working sample. The grey dots correspond to the
GAMA/AAT spectra. The orange grids represent the regions that define
the $4\times 3$ stacked spectra. We define a {\sl local} parameter,
defined by the stellar mass of the secondary galaxy, and an {\sl
environment} parameter defined either by the mass of the most massive
companion (called the primary galaxy, left panel), by the
luminosity-derived group mass (middle panel), or the dynamical group
mass (right panel). Group masses are retrieved from the G$^3$C
catalogue of \citet{G3C}, and the luminosity-derived estimates follow
the scaling relation of \citet{Viola:15}.  For reference, the blue
dashed lines on the left panel show the loci for a 1:1 and a 1:10
merger progenitor.}
\label{fig:grid}
\vskip-0.3truecm
\end{figure*}

\section{Data}
\label{Sec:Data}

We retrieve our sample from the GAMA-II database, a panchromatic
galaxy survey providing a set of spectroscopic redshifts down to
$r_{\rm AB}$=19.8\,mag \citep{Liske:15}. We focus on the equatorial
fields, that cover $\sim$180\,deg$^2$ in three regions, with a high
($\sim$98.5\%) spatially uniform redshift completeness that makes it
optimal for studies of environment \citep[see,
e.g.,][]{G3C,Brough:13}. We note that in GAMA the same fields were
repeatedly visited, so that, by construction, the spectroscopic
completeness is very high, not only in general, but also over small
scales, avoiding the standard issues found in SDSS spectroscopic data
sets regarding fibre collision.  The tiling and observing strategies
of the survey are discussed in detail in \citet{Robotham:10}
and \citet{GAMA}.

Our selection starts with the general set of
massive galaxies, defined as those with a stellar mass above
$10^{11}$M$_\odot$.  The sample is extracted from the latest version
(v18) of the catalogue of stellar masses in the GAMA
survey \citep{Taylor:11}, and restricted to the
0.1$\leq$z$\leq$0.3 redshift range, in order to minimise
aperture effects. The set comprises 12,616 massive galaxies\footnote{From
which 8,186 sources have AAT spectra, and 4,313 have SDSS spectra. The
remaining 117 spectra were compiled from other surveys (2dFGRS,
WiggleZ, etc).}. Within this sample of massive galaxies we look for
dynamically close pairs, which serve as potential merger
progenitors. A close pair is defined here by a system separated by a
projected distance less than 100\,kpc, and with a
velocity difference below 700\,\kms. Hereafter, we refer to the most
massive galaxy in the pair as a primary, and the companion is termed
either a secondary, or a satellite. This criterion yields a total of
3,770 satellites, in 2,787 systems (note that some of the primary
galaxies may have more than one satellite). From these, only 227
satellite galaxies have SDSS spectra -- as expected for such close
pair systems, whereas AAT data are available for 3,506 satellites. We
want to minimise potential biases from systematics related to the use
of spectra from different instruments.  Given the higher completeness
of the AAT spectra, we decide to use only these data in the
analysis. Furthermore, we remove low quality data from the sample,
discarding all spectra with a low value of n$_{\rm Q}$
\cite[$<$3, as defined in][]{GAMA}, or a low S/N ($<$3, as defined by
the {\tt runz} code). Moreover, a small fraction
($\sim$2\%) of the GAMA/AAT spectra have severe
fringing \citep{GAMA:Spec}.  Therefore we also remove those visually
inspected to feature such fringing, resulting in a final working sample
of 2,692 satellite spectra. The same criteria
applied to the selection of massive galaxies yields a sample of
7,702 systems.

\subsection{Sample selection effects}
\label{Ssec:apert}

The GAMA/AAT spectra were acquired through optical fibres that map
onto a 2\,arcsec diametre aperture \citep{GAMA:Spec}. Within the
redshift range of the sample ($0.1\leq z\leq 0.3$), the fibre aperture
extends over a projected physical distance between 3.7 and
8.9\,kpc. Fig.~\ref{fig:RenSz} shows the distribution of our satellite
galaxies with respect to the fibre aperture (measured in units of the
effective radius, {\sl top}); the S\'ersic index ({\sl middle}) and
the redshift ({\sl bottom}). The figure makes use of the surface
brightness fits to the SDSS imaging of the GAMA fields
from \citet{Kelvin:12}.  The results are colour-coded, to easily
detect any potential bias related to aperture effects or redshift. The
diagrams show the typical trends expected with the stellar mass of the
satellite (i.e. a horizontal colour gradient in this figure).
The lack of galaxies in the bottom-right corner of each panel
is caused by the fact that, by construction, M$_{\rm PRI}>$M$_{\rm SAT}$. Note
lower mass galaxies dominate the sample at low redshift and at lower
S\'ersic indices.  Regarding the size of the galaxy with respect to
the fibre size, there is a competing effect between the small {\sl
intrinsic} sizes of lower mass galaxies, and the small {\sl relative}
sizes of the more massive galaxies, preferentially located at higher
redshift. This explains why in the top panel of Fig.~\ref{fig:RenSz},
the largest values of relative size occur at stellar masses around
$\rm \log (M_s/10^{11}M_\odot)\sim-0.5$.  In any case, note that most
of the spectra enclose the light within, at least, an effective
radius. Therefore, the variations found in the spectra are expected to
map the general population in these galaxies, and not potential radial
gradients. Furthermore, note that in this paper we focus on a
differential analysis between satellite galaxies, at fixed stellar
mass, with respect to either the mass of the primary or the mass of
the halo where the system is located.  We conclude that no significant
systematic trend is expected from the sample selection or the use of
optical fibres.

Fig.~\ref{fig:vDplot} shows the distribution of satellite galaxies
with respect to their relative velocity ($\Delta$v$_{\rm PEC}$) and
projected separation ($\Delta$R$_\perp$). The whole sample is shown as
filled dots. In particular, red and blue dots correspond to galaxies
with the highest (M$_c>2\times 10^{11}$M$_\odot$) and lowest
($10^{11}$M$_\odot<$M$_c<1.5\times 10^{11}$M$_\odot$) values of the
mass of the primary, respectively. No systematic differences are found
in this diagram, although we note a steep decrease in the number of
targets with $\Delta$R$_\perp<$10\,kpc (horizontal dashed line).
Therefore we remove those satellites from the analysis of the general sample. In
addition to the sub samples segregated with respect to primary mass,
we also consider an additional subset made up of very close pairs, as
shown by the grey shaded area in the figure. Pairs in this region
($|\Delta v_{\rm PEC}|<300$\,km\,s$^{-1}$;
$0<\Delta$R$_\perp<50$\,kpc) provide a more direct
representation of merging progenitors.

\section{Preparing the sample for analysis}
\label{Sec:Prepare}

\begin{figure}
\begin{center}
\includegraphics[width=8.8cm]{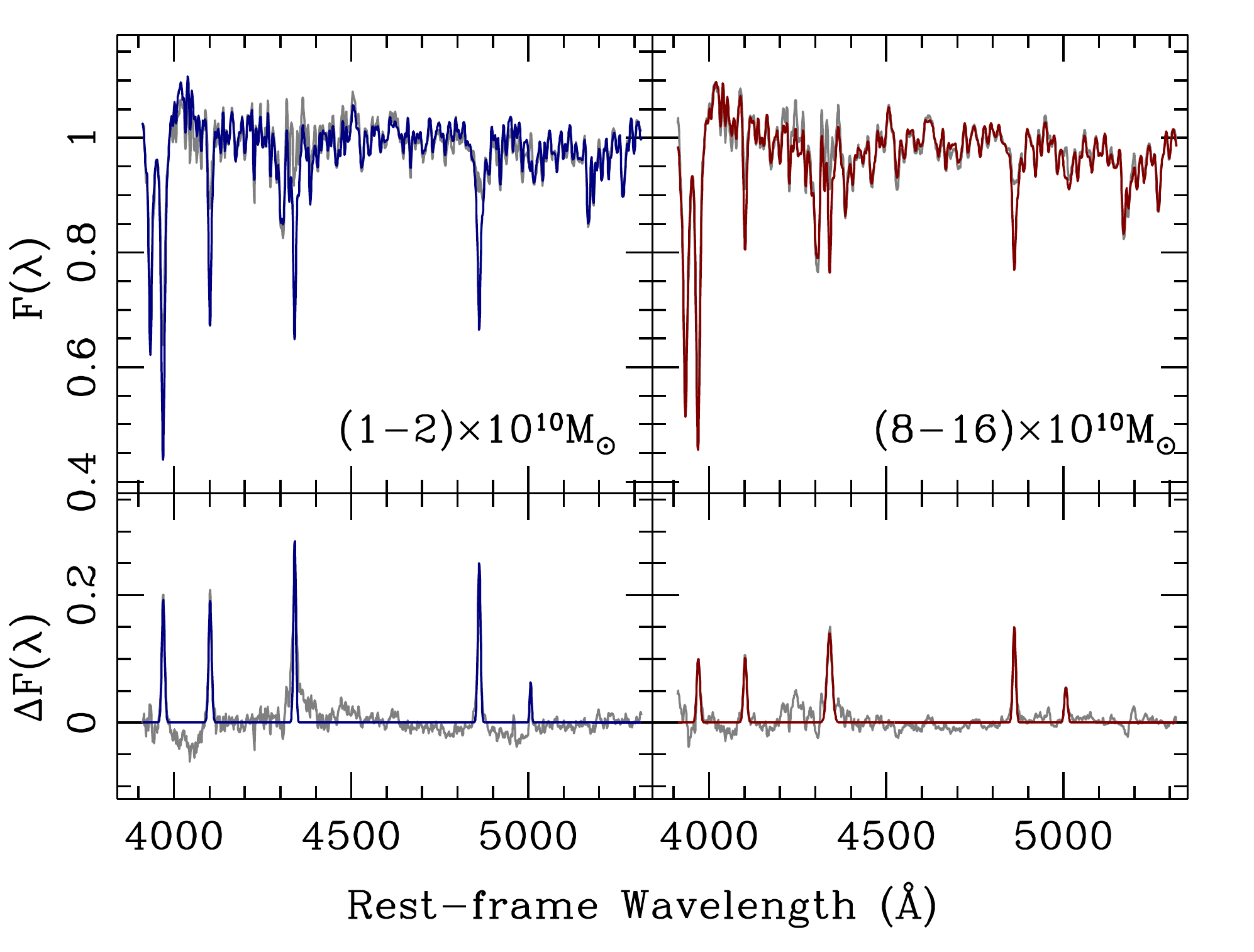}
\end{center}
\caption{Illustration of the emission line correction of
the stacked spectra. The left (right) panels correspond
to the stacked spectra of satellite galaxies with the lowest
(highest) stellar mass. The top panels plot the original
stack (in grey) and the best-fit model (in blue/red colour).
The bottom panels show the difference between the original
and the best-fit model (in grey) and between the original
and the final, cleaned spectra (in blue/red colour).
}
\label{fig:Balmer}
\end{figure}

\subsection{Individual spectra}
\label{SSec:Indiv}

Individual spectra are retrieved from the GAMA-II database and are
corrected for foreground extinction using the colour excess maps
of \citet{Schlafly:11}, following a standard dust extinction law for
the Milky Way \citep{CCM:89}. Given the typically low values of
extinction on the footprint of the GAMA survey, this step is only --
mildly -- relevant for the derivation of the 4000\AA\ break,
whereas the rest of the analysis is based on a
continuum-subtracted spectrum. The GAMA/AAT spectra present some
flux calibration issues, most notably a variable level of fringing,
and scattered light in the blue arm
\citep[see][for details]{GAMA:Spec}. We performed a number of
tests involving spectral fitting that gave us a complex range of
residuals in the spectra that were not trivial to eliminate based on
simple prescriptions.  Therefore, in order to avoid any systematics
related to flux calibration residuals, we restrict the analysis of
stellar populations to absorption line features, performing a careful
subtraction of the continuum, effectively removing any flux
calibration problem.

The pseudo-continuum is defined following a robust method laid out
in \citet{BMC}.  In a nutshell, the pseudo-continuum is defined as a
high-order percentile of the flux density values within a kernel
window. At the resolution and sampling of SDSS spectra (similar, but
not identical to GAMA data), \citet{BMC} concluded that the continuum
was best fit with a 90\% level within a 100\AA\ window. This choice is
not critical.  We note that similar values were found independently
for SDSS spectra of stars \citep{Hawkins:14}. In this case, we follow
a 3-step approach to derive the pseudo-continuum, with an initial pass
with a 100\AA\ window at the 90\% level, followed by a second pass on
the derived continuum, with a 200\AA\ window and a 90\% level. This
pseudocontinuum is subtracted from the observed spectrum. A final,
third step, fits the derived continuum with a third order polynomial
that removes any residual changes over large scales in wavelength
space. Such a technique has already been successfully applied to
GAMA/AAT spectra \citep{Baldry:14}.  All the spectra are normalized to
the same flux within the [4400,4800]\AA\ wavelength range in the
rest-frame.

In addition, we include in the analysis the strength of the 4000\AA\ break,
D$_n$(4000), as defined by \citet{Dn4000}. This feature
-- measured prior to continuum removal -- extends over a 250\AA\
region, and so, we expect the results not to be significantly
affected by flux calibration issues. In contrast to the other
line strengths, this index is measured on individual
spectra, and the values corresponding to each set of stacked spectra
are combined to provide the average and uncertainty of the index of
a given stack.

\begin{figure}
\begin{center}
\includegraphics[width=8.7cm]{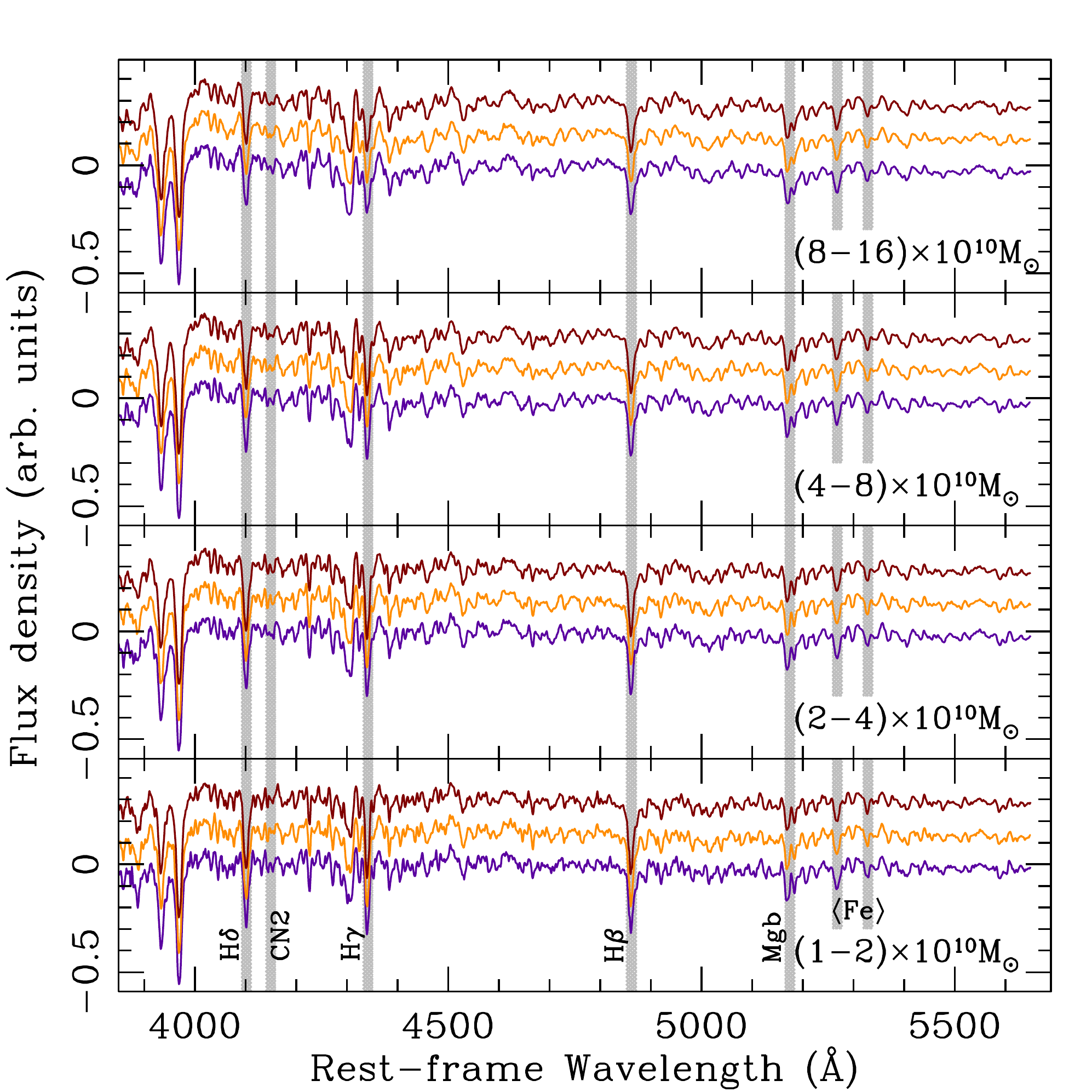}
\end{center}
\caption{Continuum-subtracted stacked spectra of the satellite galaxies.
From bottom to top, the different panels correspond to increasing
values of the stellar mass of the satellite, as labelled.  Within each
panel, three stacks are shown, in red, orange and blue, corresponding
to decreasing values of the stellar mass of the primary -- as shown in
the grids of Fig.~\ref{fig:grid}. The spectra have been corrected for
emission in the Balmer lines (see text for details). For ease of
visualization, the spectra have been shifted arbitrarily along the
vertical direction (flux).  The grey shaded areas mark typical
spectral features used in the analysis, from blue to red: H$\delta$,
CN2, H$\gamma$, H$\beta$, Mgb, Fe5270 and Fe5335.}
\label{fig:stacks}
\end{figure}

\begin{figure*}
\begin{center}
\includegraphics[width=13.5cm]{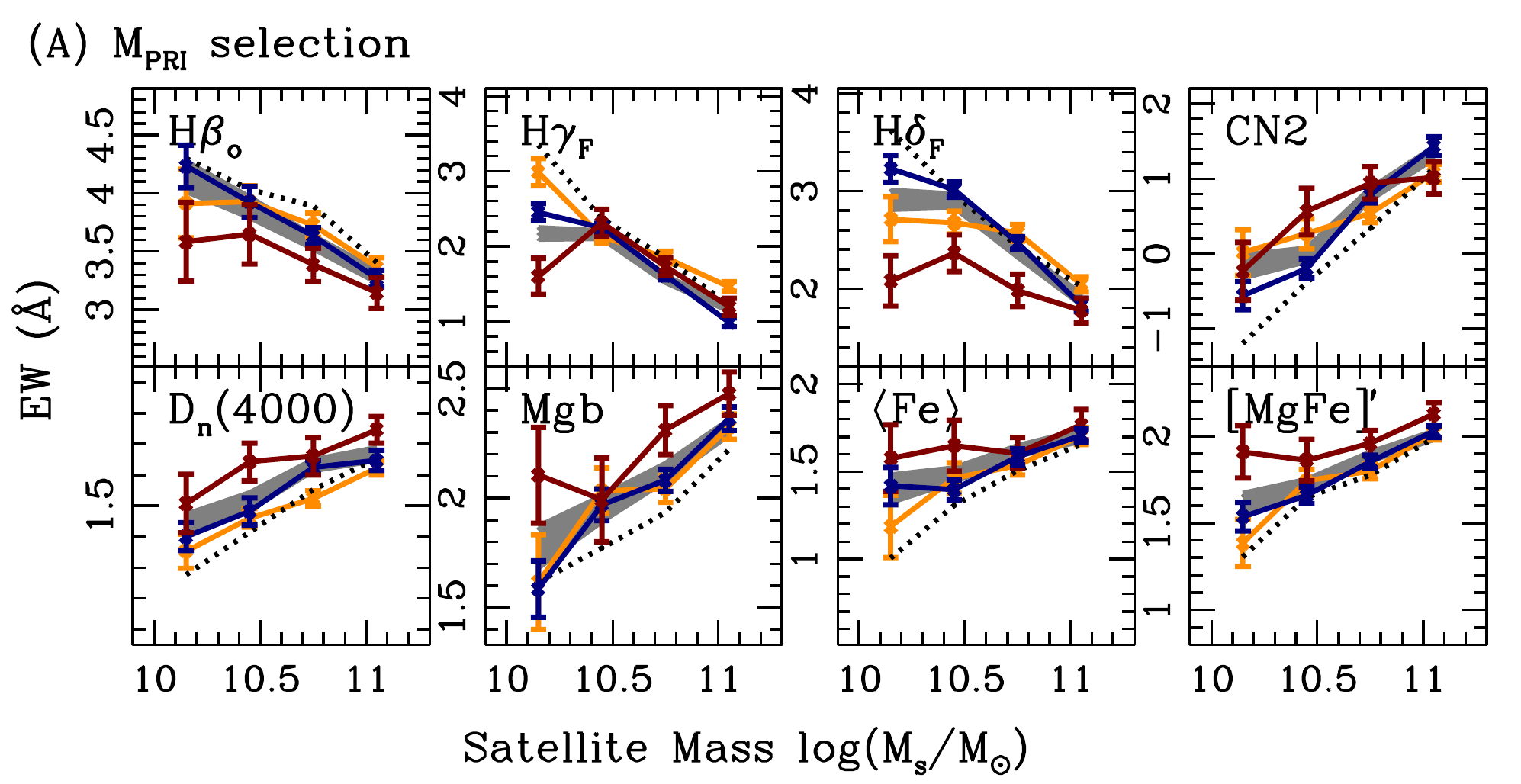}
\includegraphics[width=13.5cm]{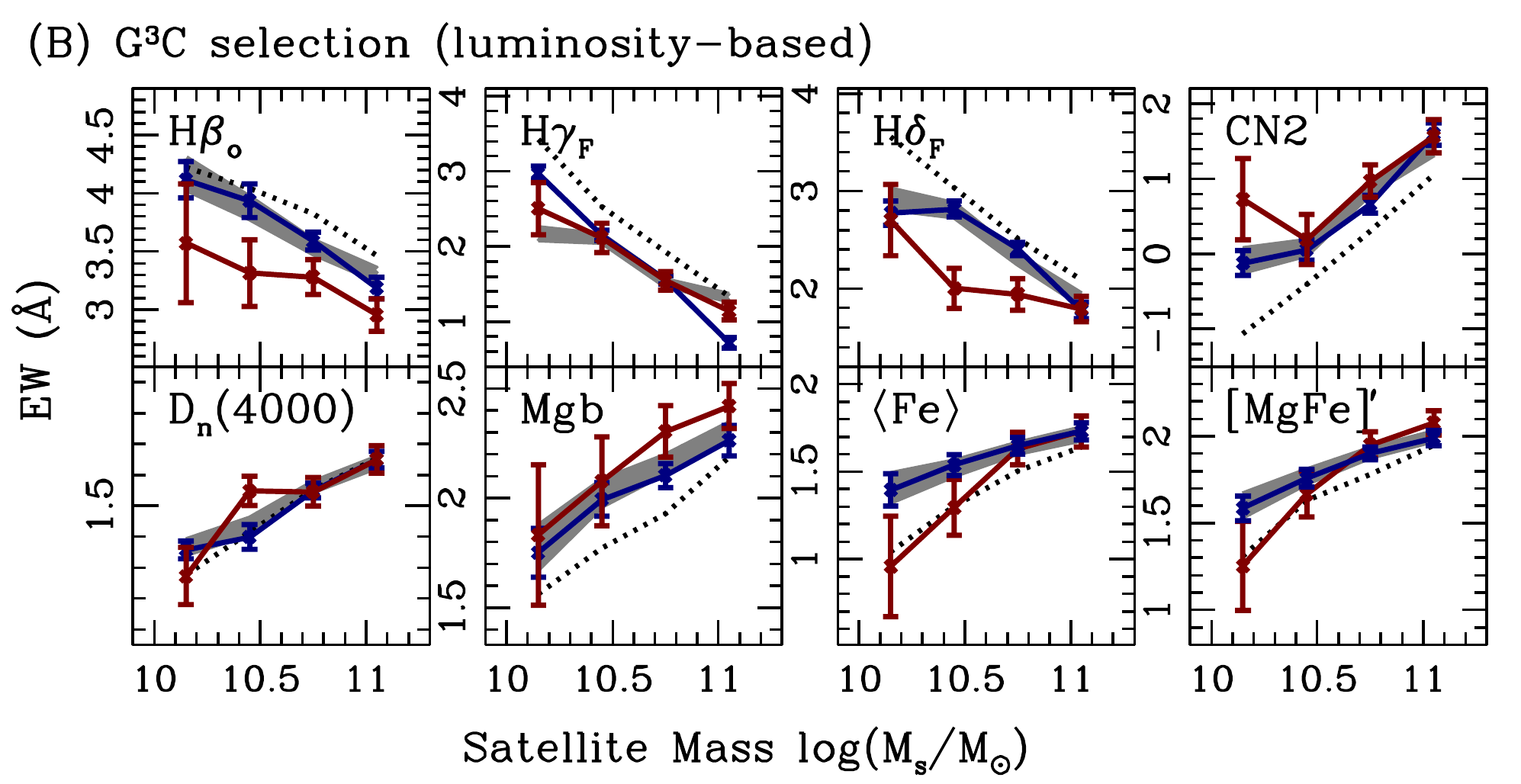}
\includegraphics[width=13.5cm]{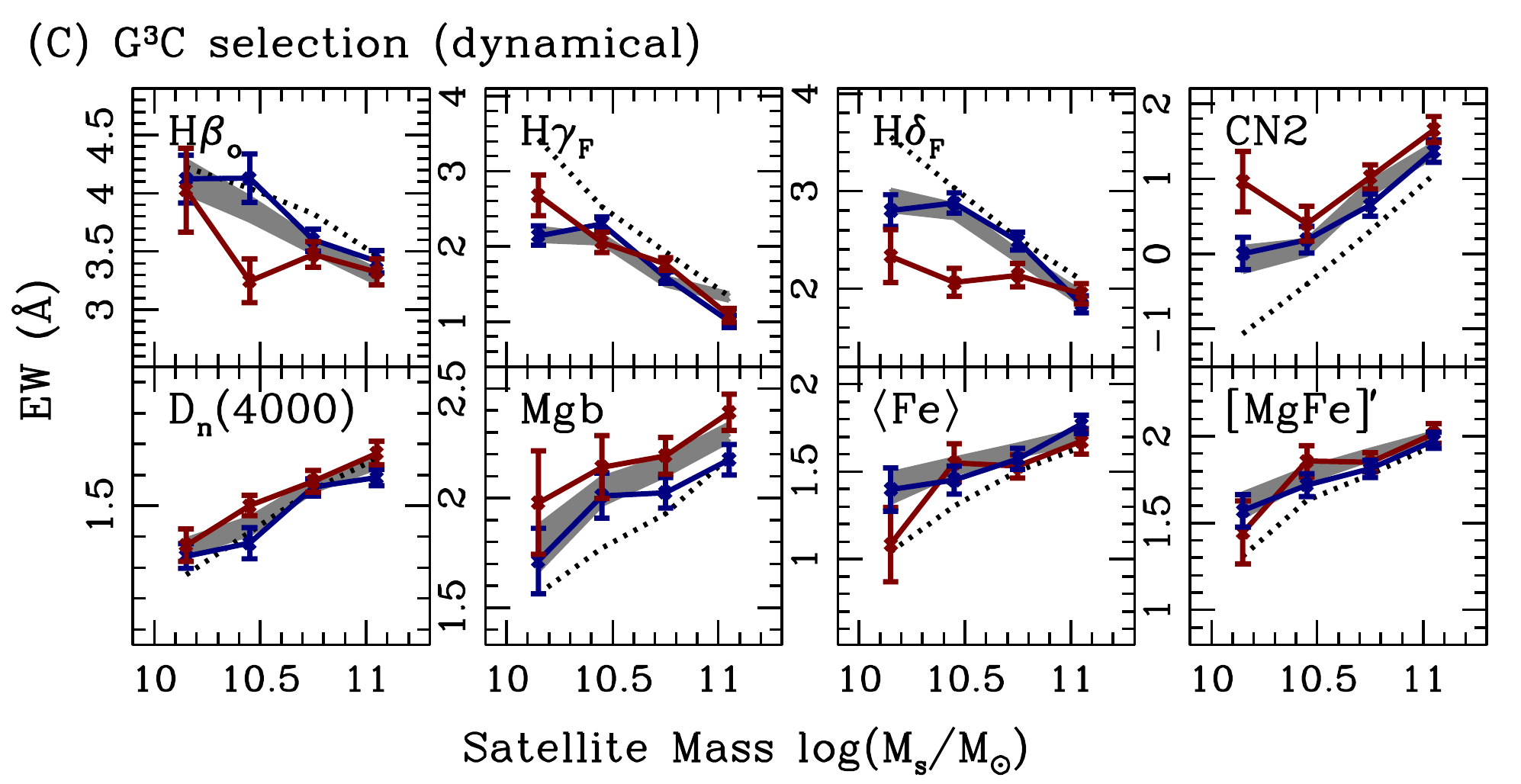}
\end{center}
\caption{Equivalent widths of stacked spectra of satellite galaxies
in close pairs, shown as a function of the stellar mass of the galaxy,
and a second, environment-related parameter, namely the mass of the
primary galaxy (A, top panels) or the group mass (B,C, middle and bottom panels)
according to the G$^3$C catalogue of \citet{G3C}. The grey shaded area
is the trend for stacks made irrespective of the environment
parameter. The blue (red) data points are the values for satellites
around the least (most) massive primary galaxy, or group mass,
corresponding to the bottom and top rows in the grids shown in
Fig.~\ref{fig:grid}. The orange points in the top panels represent a
subsample comprised of systems where the closeness criterion is
tighter (see text for details).  The dotted black line corresponds to
the large sample from all GAMA spectra (i.e. stacked with the same
criterion as with the satellite galaxies regarding S/N, but
irrespective of whether they are located in close pairs).
The halo masses in
(B) are derived from the total group luminosity, whereas in (C) the
masses are derived from a dynamical argument (see text for details).}
\label{fig:EWs}
\end{figure*}

\begin{figure}
\begin{center}
\includegraphics[width=8.5cm]{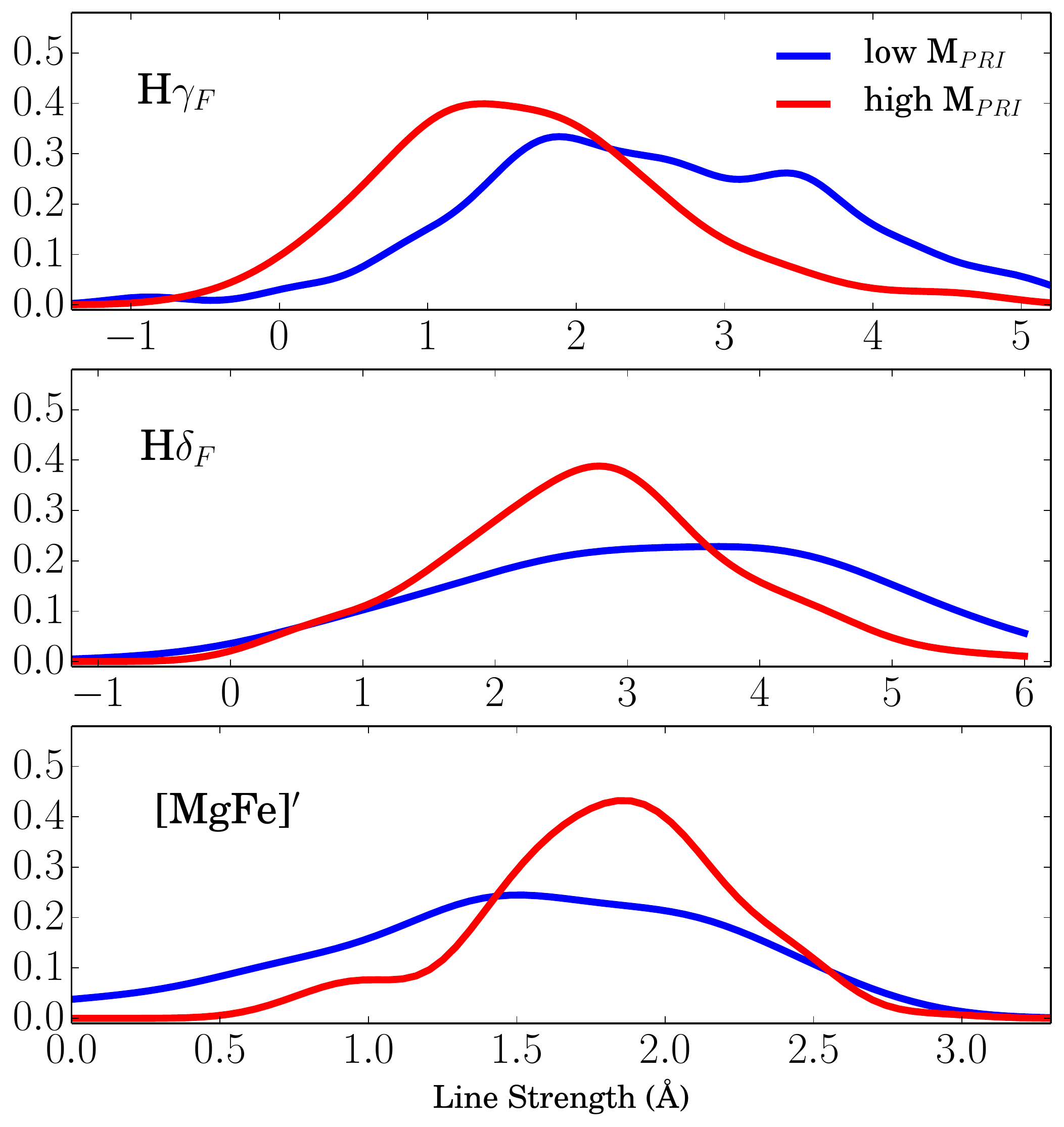}
\end{center}
\caption{Distribution of equivalent widths measured in
{\sl individual} spectra. They correspond to satellite galaxies in the
lowest stellar mass bin, stacked according to primary mass (shown in
red/blue, as labelled).  They show that the differences found in the
stacked measurements are representative of the whole sample, and are
not affected by a few spectra with higher S/N or deeper line
strengths.}
\label{fig:IndivEWs}
\end{figure}

\begin{figure*}
\begin{center}
\includegraphics[width=8.8cm]{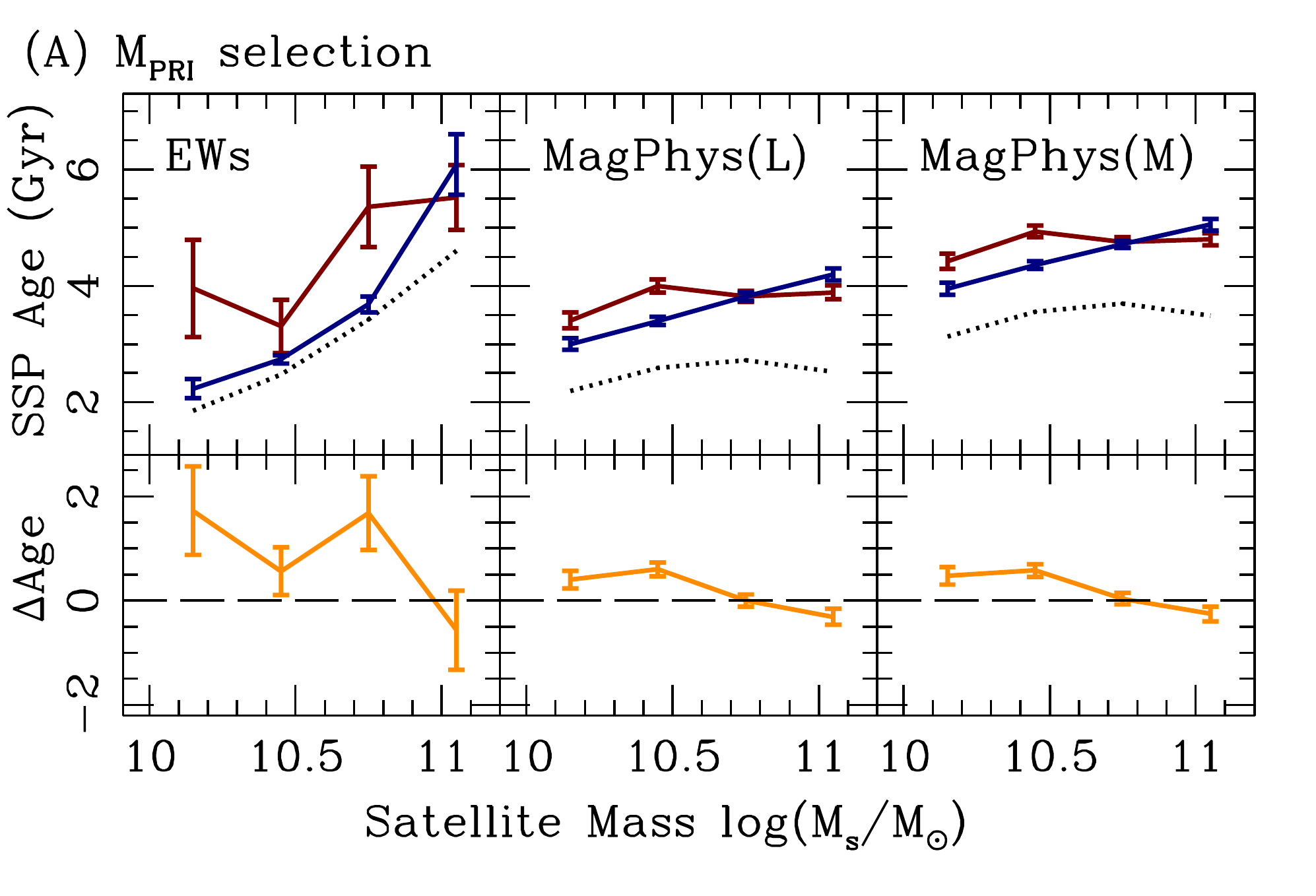}
\includegraphics[width=8.8cm]{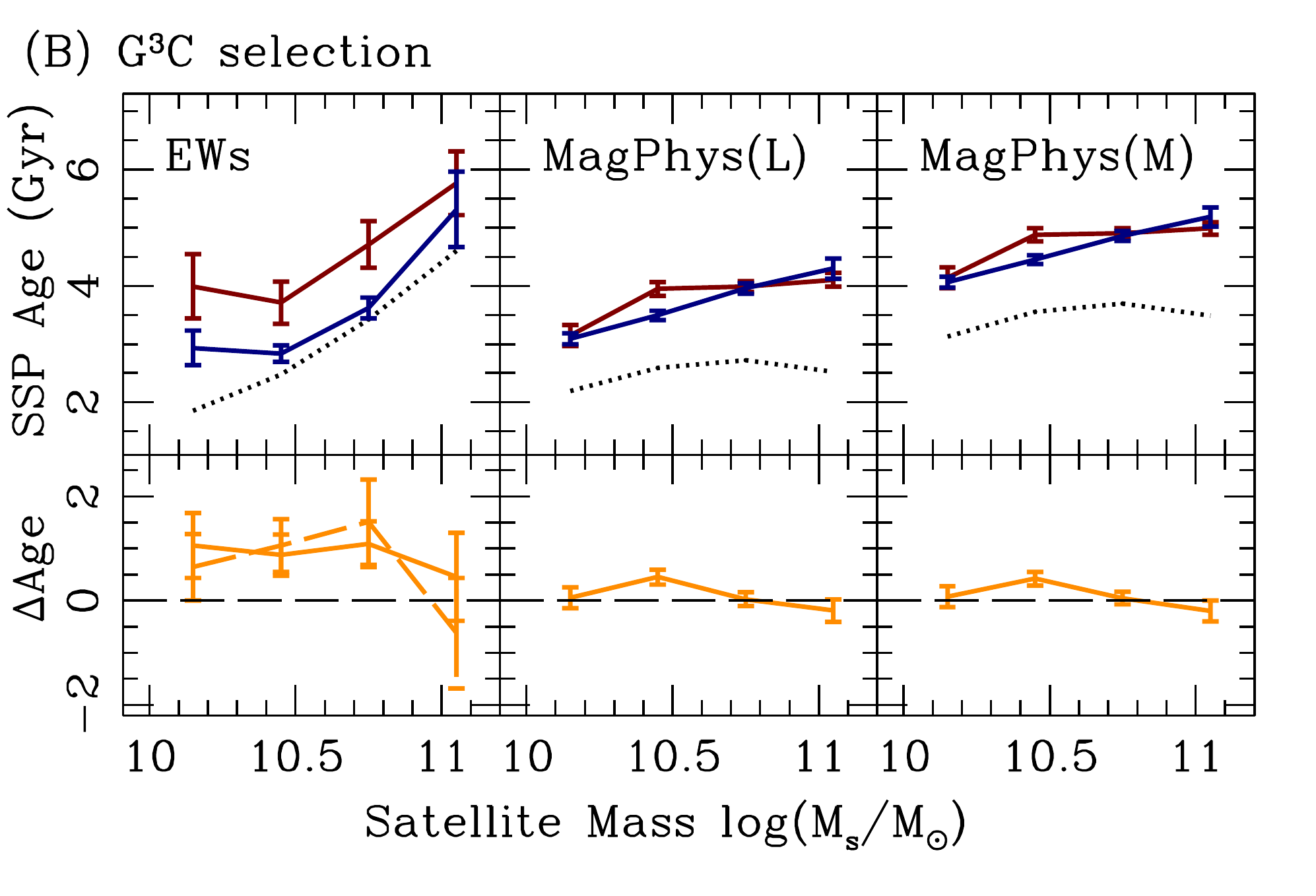}
\end{center}
\caption{Age comparison from the line strength analysis (left) and 
 MagPhys (middle and right panels). The blue/red colour coding is the
same as in Fig.~\ref{fig:EWs}. The dotted black line corresponds to
the large sample comprising all high S/N GAMA spectra within the
same redshift window. The orange lines in the bottom panels
show the age difference between the red and the blue lines in the
top panels, i.e. between satellites next to the most and the
least massive primary galaxies. The leftmost panels (A) represent
the stacked data according to the stellar mass of the primary,
whereas the rightmost panels (B) correspond to stacks following 
the (luminosity-based) halo mass. We also include a dashed
orange line representing the age difference when binning with
respect to the dynamical halo mass estimates.}
\label{fig:Age}
\end{figure*}

\subsection{Stacking procedure}
\label{SSec:Stack}

Our analysis is based on spectral line strengths to constrain the age
and chemical composition of the populations. Such an approach requires
relatively high S/N, leading us to stack spectra, discarding the very
noisy ones. Our final working sample only includes spectra with S/N
(as defined by the {\tt runz} code) above 3.  We illustrate the
selection of individual spectra for stacking in Fig.~\ref{fig:grid},
where the orange lines delimit the binning scheme.  We use the stellar
mass of the secondary as the ``local'' parameter (horizontal axis).
The ``environment'' parameter is defined either as the mass of the
primary (left panel), or the mass of the group within which the pair
live (middle and right panels). The groups are defined in the G$^3$C
catalogue of \citet[][we use v09]{G3C}, and are translated into halo
masses following either the scaling relation with total luminosity
from \citet{Viola:15}, or the dynamically-based halo masses
of \citet{G3C}, from derived estimates of the size and velocity
dispersion of the group. For the latter, we use their mass proxy
corrected by the $A$ factor. The quantity $\mu\equiv M_{\rm SAT}/M_{\rm PRI}$
can be interpreted as the mass ratio of the eventual merging system.

The grey dots in Fig.~\ref{fig:grid} correspond to secondary galaxies
with available GAMA/AAT spectra. As reference, two values of the mass
ratio (1:1 and 1:10, corresponding to $\mu=1$ and $0.1$, respectively)
are shown as blue dashed lines on the left
panel. Tab.~\ref{tab:stacks} shows the number of spectra used for the
analysis presented in this paper. Note that not all of the
galaxies from the close pair sample are listed in the groups catalogue.
In addition, we consider two
additional samples, a subset of ``very close'' pairs -- as defined by
the shaded region in Fig.~\ref{fig:vDplot} -- consisting of systems
with a small relative velocity and projected separation, and a large
sample made up of all GAMA/AAT spectra with the same constraints with
respect to S/N, redshift, etc, but regardless of whether they are
close to a massive galaxy.  This sample will help to compare the
properties of the satellite galaxies with respect to the general
population. This larger sample is hereafter termed the field sample.
Tab.~\ref{tab:stacks2} shows the number and S/N of these
two additional sets.

After following the steps described in \S\S\ref{SSec:Indiv} for the
individual spectra, the continuum-subtracted data are stacked, with a
weighting scheme following a sigmoid function with a threshold at SNR=8,
namely
\begin{equation*}
\rm w(SNR) = 1- \frac{1}{e^{SNR-8}+1},
\end{equation*}
where SNR is the average SNR per \AA\ within the rest-frame interval
4000--4300\AA. This scheme weighs equally all data points within the
same individual spectrum. We note that the standard scheme, using the
inverse variance as a weight, gives slightly noisier stacks.  The
resulting spectra are corrected for nebular emission. We perform
spectral fitting using the code {\tt STARLIGHT} \citep{Starlight},
using as basis functions a set of 92 model spectra
from \cite{MIUSCAT}, after applying the same continuum normalization
as on the observed spectra. The basis functions cover a wide range of
ages -- from 0.3 to 13.7\,Gyr in 23 steps in logarithmic space, and
metallicity -- from $\log(Z/Z_\odot)=-0.5$ to $+0.2$ in 4 steps,
logarithmic with respect to Z. The output gives the effective velocity
dispersion of the stack, and a best fit spectrum.
Fig.~\ref{fig:Balmer} shows the procedure in two  stacks,
where the difference between the observed spectra and the best-fit
model (bottom panels) reveals the presence of the emission lines.
These lines (H$\beta$,H$\gamma$,H$\delta$,[O{\sc III}]) are fit in the
residual, using Gaussian profiles, following a Levenberg-Marquardt
algorithm. These fitted lines are removed from the original stack
and the final, emission-corrected spectrum is smoothed to a fiducial
velocity dispersion of 200\,km/s, using a Gaussian kernel.  The line
strengths are measured on the final spectra, using the uncertainties
in the stacks to bootstrap the errors on the equivalent widths from an
ensemble of 100 realisations.

\begin{table*}

\caption{Number and S/N (in brackets) of spectra used in the stacks
(see Fig.~\ref{fig:grid}). The S/N is given per pixel, averaged in the
region around the Mgb feature ($\lambda\sim$5175\AA).  The group halo mass
is derived either from the total luminosity given by the G$^3$C catalogue
of \citet{G3C}, following the scaling of \citet{Viola:15},
or by the dynamical estimate presented in \citet{G3C}, as labelled.
\label{tab:stacks}}
\begin{tabular}{c|cccc}
\hline
 & \multicolumn{4}{c}{log M$_{\rm PRI}$/M$_\odot$ stacks}\\
\hline
log M$_{\rm SAT}$/M$_\odot$ & 11.0-11.2 & 11.2-11.4 & 11.4-11.6 & ALL \\
\hline
10.0-10.3 & 217 ( 72) & 164 ( 57) &  58 ( 45) & 439 (102)\\
10.3-10.6 & 391 (118) & 254 ( 92) & 114 ( 75) & 759 (167)\\
10.6-10.9 & 440 (144) & 262 (116) & 132 (101) & 834 (211)\\
10.9-11.2 & 215 (122) & 171 (134) &  74 (100) & 460 (207)\\
\hline
 & \multicolumn{4}{c}{log M$_{\rm GRP,L}$/M$_\odot$ stacks: Luminosity-based}\\
\hline
log M$_{\rm SAT}$/M$_\odot$ & 13.2-13.7 & 13.7-14.2 & 14.2-14.7 & ALL \\
\hline
10.0-10.3 & 247 ( 61) & 119 ( 34) &  35 ( 19) & 401 ( 72)\\
10.3-10.6 & 371 ( 96) & 258 ( 73) &  74 ( 38) & 703 (127)\\
10.6-10.9 & 349 (124) & 295 (104) & 114 ( 62) & 758 (173)\\
10.9-11.2 & 124 (102) & 218 (135) &  69 ( 70) & 411 (183)\\
\hline
 & \multicolumn{4}{c}{log M$_{\rm GRP,D}$/M$_\odot$ stacks: Dynamical}\\
\hline
log M$_{\rm SAT}$/M$_\odot$ & 12.50-13.25 & 13.25-14.00 & 14.00-14.75 & ALL \\
\hline
10.0-10.3 & 142 ( 46) & 176 ( 49) &  83 ( 25) & 401 ( 72)\\
10.3-10.6 & 229 ( 73) & 308 ( 90) & 169 ( 54) & 706 (127)\\
10.6-10.9 & 242 ( 98) & 317 (115) & 203 ( 87) & 762 (174)\\
10.9-11.2 & 112 (102) & 194 (125) & 102 ( 85) & 408 (183)\\
\hline
\end{tabular}
\medskip
\end{table*}

\begin{table}
\caption{Number and S/N (in brackets)
of additional spectra used here. The S/N is computed on the stacked data.
\label{tab:stacks2}}
\begin{center}
\begin{tabular}{cc}
\hline
log M$_{\rm SAT}$/M$_\odot$ & Number (S/N)\\
\hline
\multicolumn{2}{c}{Close sample$^1$}\\
\hline
10.0-10.3 & 108 ( 48)\\
10.3-10.6 & 212 ( 89)\\
10.6-10.9 & 238 (121)\\
10.9-11.2 & 181 (133)\\
\hline
\multicolumn{2}{c}{Field sample$^2$}\\
\hline
10.0-10.3 & 14631 (1409)\\
10.3-10.6 & 21705 (2010)\\
10.6-10.9 & 22044 (2403)\\
10.9-11.2 & 10659 (2162)\\
\hline
\end{tabular}
\end{center}

\medskip

$^1$ Defined by $\Delta v_{\rm PEC}<$300\,km\,s$^{-1}$, $\Delta R_\perp<$50\,kpc.\\
$^2$ Comprises all GAMA/AAT spectra with the same constraints as in the close 
pair sample, except for the proximity to a massive galaxy.
\end{table}


\section{Methodology}
\label{Sec:Method}

Fig.~\ref{fig:stacks} shows the continuum-subtracted stacks from our
sample when the ``environment'' parameter is defined as the stellar
mass of the primary. From bottom to top, the sample is shown in
increasing stellar mass of the satellite, as labelled, whereas in each
panel three spectra are shown, corresponding from bottom to top to an
increased mass of the primary (see grids on the left-hand panel of
Fig.~\ref{fig:grid} for reference). The spectra are shifted
vertically by a constant amount for clarity. The vertical shaded
regions encompass the line strengths used in the analysis.  Note the
significant variation in effective resolution with respect to
satellite mass, caused by the velocity dispersion of the stars,
especially evident in the Mgb feature at $\lambda\sim$5170\AA. This
trend is removed from the analysis by smoothing all the spectra to a
common fiducial value of 200\,km\,s$^{-1}$.

The continuum-subtracted stacked spectra provide information
about the underlying stellar populations of the putative
progenitors of massive galaxies. We can therefore explore
differences in targeted line strengths to probe the
characteristics of the populations that will be eventually
incorporated in massive galaxies at later times.
Although the details of the process vary substantially from system
to system, we can generally assume that in minor merging
systems, the populations of the satellite galaxy will be
incorporated in the outer envelope of the merged system
\citep[see, e.g.,][]{Naab:09}, whereas a major merger will produce a more
efficient spatial mixing of both galaxies.

The analysis is performed by comparing the observed, stacked data with
population synthesis models. We choose the latest version of the
MIUSCAT models from \citet{MIUSCAT}. These models provide the
spectra of simple stellar populations at a 2.51\AA\ resolution
\citep[FWHM,][]{MILES:Res}, over the $\lambda\lambda$3465-9469\AA\
spectral window for a range of ages, metallicities and stellar initial
mass functions (IMF). We note that in our analysis (restricted to
rest-frame wavelengths bluer than 5,500\AA) the MIUSCAT models are
fully based on the MILES stellar library \citep{MILES}.  For
simplicity, we adopt the standard Kroupa-universal
IMF \citep{KuIMF}. The synthetic spectra are processed following the
identical methodology as the GAMA/AAT spectra, after being convolved
with a Gaussian kernel to the fiducial velocity dispersion of
200\,km\,s$^{-1}$ chosen for all the stacked satellite spectra.  The
observed (\{$o_i\pm \sigma_i$\}) and model (\{$m_i$\}) line strengths
are compared with a standard $\chi^2$ statistic:
\begin{equation*}
\chi^2(t,Z)\equiv \sum_i\left[\frac{o_i-m_i(t,Z)}{\sigma_i}\right]^2,
\end{equation*}
where the indices used in the analysis are:
\{$o_i$\}=\{H$\beta_o$,H$\gamma_F$,H$\delta_F$,CN2,D$_n$(4000),[MgFe]$^\prime$\},
comprising the standard age-sensitive Balmer lines:
H$\beta_o$ -- following the definition of \citet{Hb0}, H$\gamma_F$ and
H$\delta_F$ \citep{Hgd}. The CN2 index \citep{Trager:98} is also
included, as previous work in the literature report variations with
respect to environment \citep{Carretero:07} and could be potentially
used as a stellar clock in addition to [Mg/Fe]$^\prime$. We 
include in the analysis the 4000\AA\ break strength \citep{Dn4000}, and the standard
metallicity-sensitive indices: Mgb, $\langle$Fe$\rangle$=Fe5270+Fe5335
and [MgFe]$^\prime$ \citep{MgFep}.  We note that although these line strengths follow
the standard definitions of the index and the red/blue sidebands they
are measured on continuum-subtracted data, except for D$_n$(4000).

The model data are derived from a grid of 1024 SSP models, taking 16
steps in metallicity, from $\log(Z/Z_\odot)=-0.7$ to $+0.2$ and 64
log-steps in age, from 0.3 to 13.7\,Gyr. Note that in order to obtain
an accurate estimate of SSP-equivalent ages and metallicities, this
grid is much denser than the one used in \S\S\ref{SSec:Stack} to
obtain a best-fit spectrum to correct for the effect of emission
lines.  We do not use Mgb or $\langle$Fe$\rangle$ in the analysis, as
the combined index, [MgFe]$^\prime$ already provides the constraint on
the metallicity, independently of ${\rm [\alpha/Fe]}$ enhancement
\citep{MgFep}.
Since we are using a reduced set of measurements, we only
consider SSP-equivalent variations. Although we warn that constraints
based on SSPs need not  provide an accurate estimate of absolute ages and metallicites,
{\sl differential} variations in stellar populations are captured
quite accurately by SSP-equivalent parameters
\citep[see, e.g.][]{BMC}. Furthermore, the
added degeneracies inherent to the use of composite populations
may wash out variations in the spectral features. 

\begin{figure*}
\begin{center}
\includegraphics[width=8.8cm]{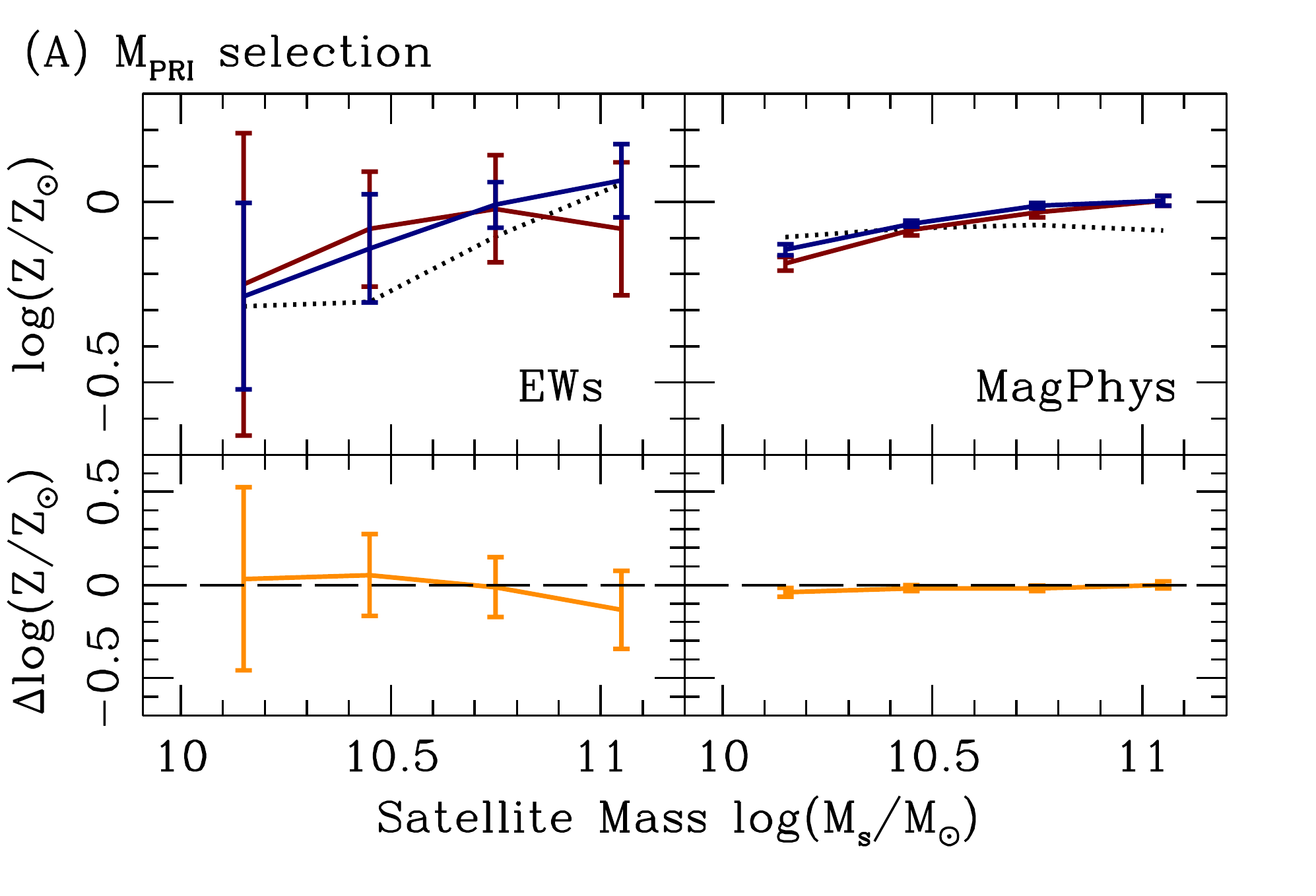}
\includegraphics[width=8.8cm]{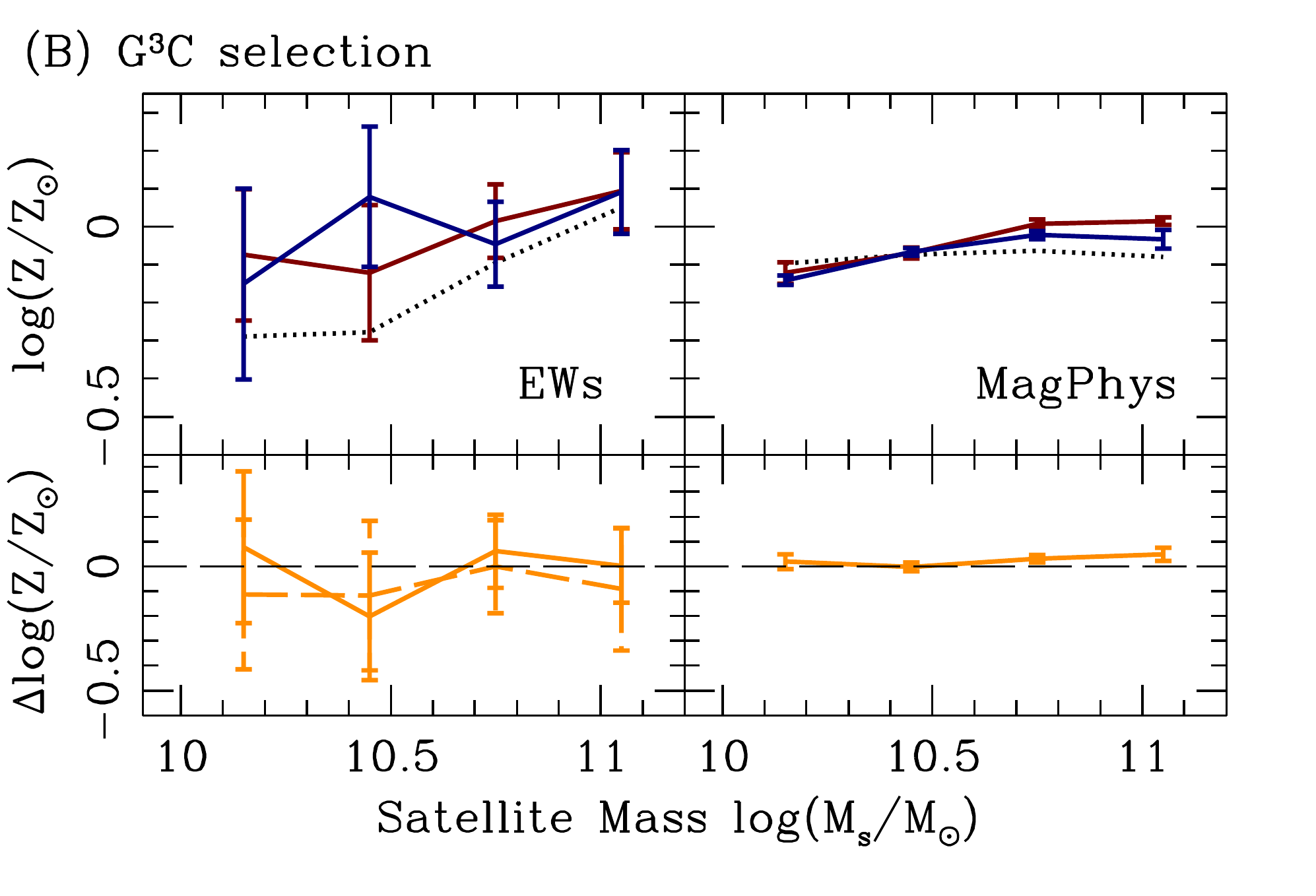}
\end{center}
\caption{Metallicity comparison from the line strength analysis 
and with MagPhys. The notation is the same as in
Fig.~\ref{fig:Age}.}
\label{fig:ZH}
\end{figure*}

\section{Discussion}
\label{Sec:Disc}

Fig.~\ref{fig:EWs} shows the trend in the equivalent widths (EWs) of
the age- and metallicity-sensitive features, as a function of the
stellar mass of the satellite galaxies. We emphasize here that the aim
is to look for differences in the properties of satellites close to
massive galaxies -- whose stellar populations will be eventually
merged. Three sets are presented corresponding to the stacking
criteria discussed above. The top panels (labelled A) show the EWs
when stacking according to the stellar mass of the primary: the red
(blue) lines correspond to satellites around the most (least)
massive primary galaxy. In addition, we show in orange the EWs for
the subsample of satellites within a closer range of the massive
galaxy (i.e. the shaded region in Fig.~\ref{fig:vDplot}). The
middle and bottom panels (labelled B and C) show the results for the 
alternative stacking procedure, based on the halo mass of the hosting
groups. In this case, the red
(blue) colours correspond to galaxies lying in the most (least)
massive groups. The difference between the middle and bottom panels lie in the 
definition of halo mass: (B) uses the luminosity-derived masses
from \citet{Viola:15}, whereas (C) uses the dynamical estimates
from \citet{G3C} (see \S\S\ref{SSec:Stack}).
In all three sets (A-C), the grey shaded regions extend over the
range of the satellite sample in general, i.e. only stacked with
respect to satellite mass, regardless of primary mass or group
mass. The dotted lines in both sets are the results for the larger
{\sl field} sample, i.e. not restricted to the presence of a close
pair, but with the same redshift distribution.

The EWs of the stacked satellite spectra reveal a significant
difference in the stellar populations {\sl at fixed stellar mass}, so
that the satellites of the most massive primary galaxies (red lines)
are slightly older and more metal rich. In addition, note that the
general sample of field galaxies (dotted lines) feature younger and
metal-poorer populations at fixed stellar mass. Therefore, there is a
significant environment-related trend where star formation proceeded
more efficiently when satellites are located in the proximity of a
more massive galaxy.  One could argue that this is a group-related
trend, so that satellites close to the most massive galaxies tend to
live in more massive halos.  Moreover, galactic
conformity \citep{Weinmann:06} poses that the properties of galaxies
within a group correlate with the properties of the central galaxy
sitting at the centre of the dark matter halo.  Therefore, we should
consider whether the observed trend is either caused by the
short-range effect of being in a close pair, or, rather, by
group-related mechanisms. The second set of panels
(B and C) -- corresponding to group-based stacking -- gives similar
trends when considering halo masses. We also note that since the
total luminosity in a group has a stronger correlation with the
luminosity of the central galaxy, it is difficult to disentangle
(central) galaxy mass and (luminosity-derived) halo mass,
whereas dynamical masses are derived from independent estimates of the
halo properties, such as size or velocity dispersion.
Of the various ways of measuring halo mass, the one
based on the correlation with total luminosity is supposed to be less
biased than the dynamical estimates \citep{Han:15}.
\citet{Hartley:15} explored galactic conformity at higher
redshift (z$\simlt$2) finding no significant evolution with cosmic
time. They concluded that a halo mass-independent mechanism could be
responsible for this trend, such as a hot halo produced by
the massive companion \citep[cf.][]{Kawinwanichakij:16}. In this
context, our trends cannot disentangle the difference between the
effect of the primary or a potential mechanism caused by the
interaction with the halo where the pair is embedded.

We need to test whether these trends are representative of the
observed sample. Fig.~\ref{fig:IndivEWs} shows the distribution of a
few line strengths measured on {\sl individual} spectra, corresponding
to satellite galaxies in the lowest stellar mass bin,
i.e. $(1-2)\times 10^{10}$M$_\odot$.  The red (blue) lines represent
the satellites around the most (least) massive primary
galaxies. Therefore, these distributions are the equivalent of the
leftmost blue and red data points in the topmost panels of
Fig.~\ref{fig:EWs}. We emphasize that the uncertainties derived from
the individual spectra are very large. The figure tests the hypothesis
that the observed trends could be an artefact of the stacking
procedure, where the signal would be originating only from a few
spectra, either because of their higher S/N or the presence of
significantly different equivalent widths. The distributions are shown
following a standard gaussian kernel density estimator (with a kernel
size $\sigma/5$, where $\sigma$ is the standard deviation of the
distribution). The individual measurements reveal the same trend as in
the stacked data, confirming the trends are robust against stacking
artefacts.

\begin{figure*}
\begin{center}
\includegraphics[width=8cm]{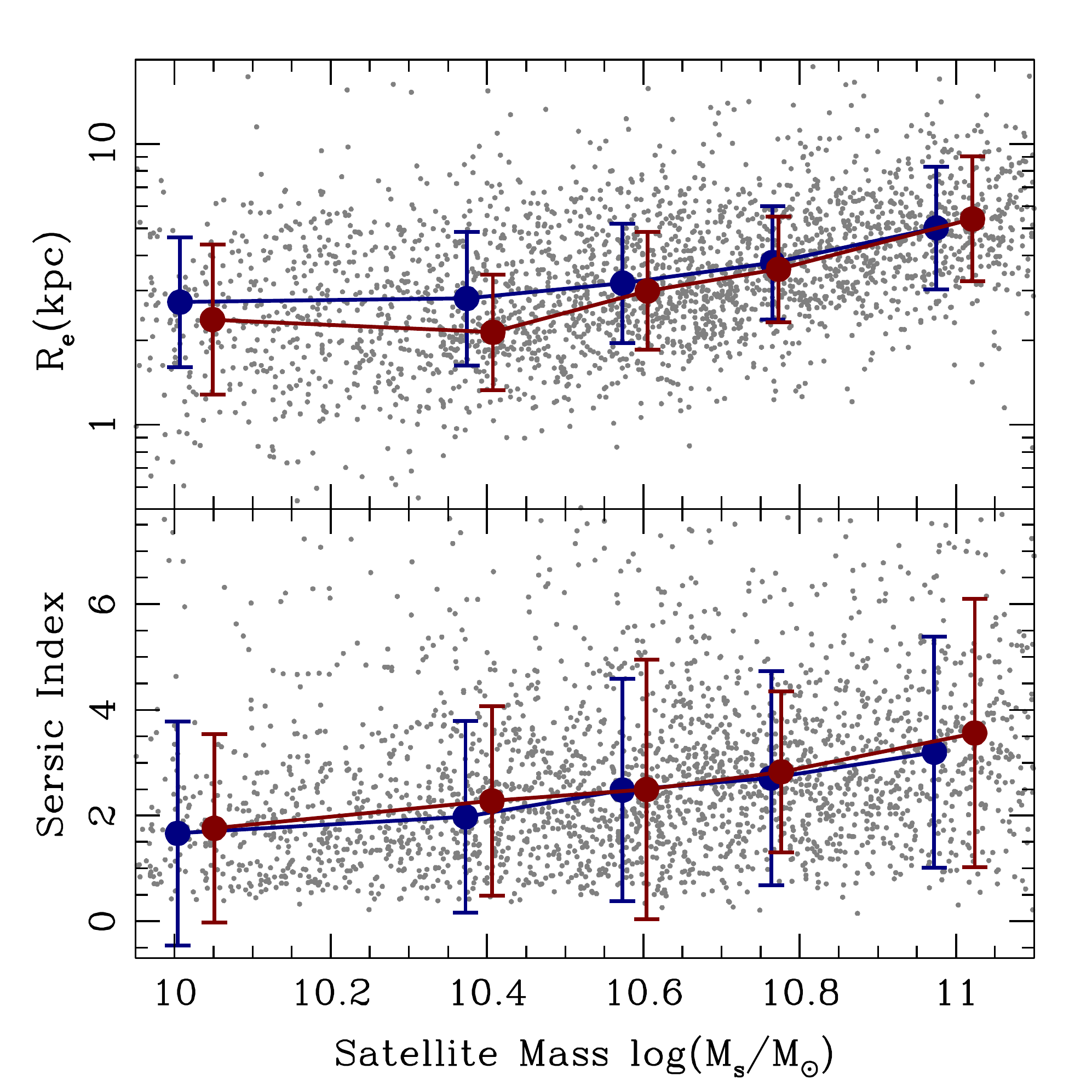}
\includegraphics[width=8.8cm]{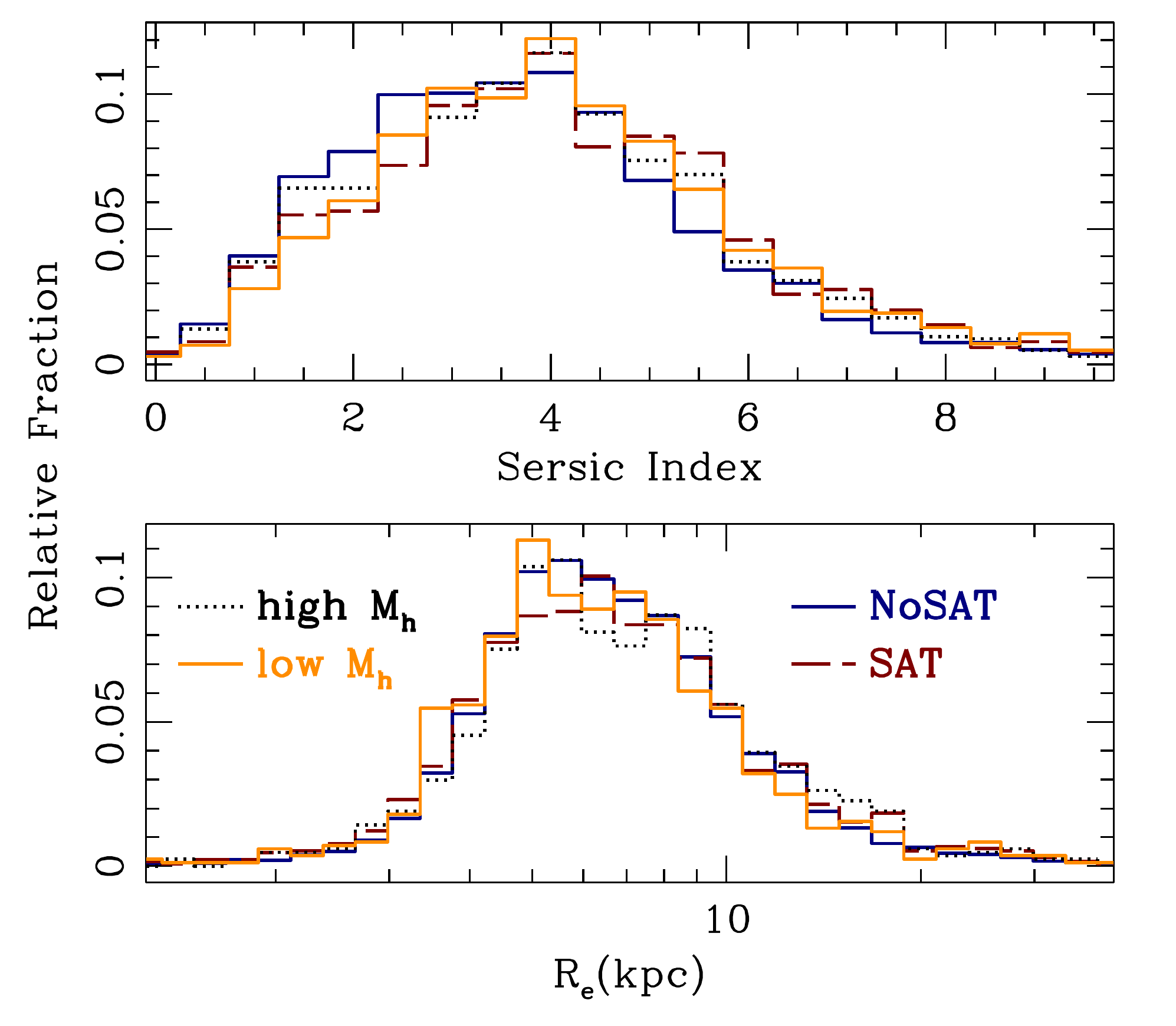}
\end{center}
\caption{Left: Distribution of structural parameters in the sample of
satellite galaxies. The effective radius (top) and S\'ersic index
(bottom) is presented as a function of the stellar mass for the sample
of satellite galaxies. Grey dots correspond to individual galaxies, whereas
the dots and error bars are averages and (RMS) scatter
for data binned at fixed number of galaxies per bin. The red (blue) symbols
correspond to satellites close to the most (least) massive primary galaxies,
i.e. the same criterion as in, e.g., Fig.~\ref{fig:EWs}.
Right: Distribution of structural parameters for the
primary galaxies, split with respect to the presence of a satellite or
in halo mass. For the latter, we only choose primaries where
the luminosity-derived halo mass lies in the lowest quartile
(low M$_h$) or the highest quartile (high M$_h$).
}
\label{fig:Sersic}
\end{figure*}

In order to translate these EW measurements into physical parameters,
we apply a simple comparison of the data with a grid of simple stellar
populations (see \S\ref{Sec:Method}). Note that since we are only
using a reduced set of observations -- the noise level and flux
calibration of the GAMA/AAT spectra prevent us from applying full
spectral fitting methods -- we restrict the analysis to SSP-equivalent
values of age (Fig.~\ref{fig:Age}) and metallicity
(Fig.~\ref{fig:ZH}). The top panels give the values along with the
1$\sigma$ uncertainty, with the same colour coding as in
Fig.~\ref{fig:EWs}. The orange lines in the bottom panels give the
difference between the two extreme cases in red and blue, shown in the
top panels. In this figure we add to our (EW-based) estimates, those
from the SED fitting analysis based on {\sc magphys} \citep{MAGPHYS},
where Fig.~\ref{fig:Age} includes the constraint on the mass- (M) and
SDSS-$r$ band luminosity-weighted (L) age. The {\sc magphys}
and the EW-based constraints originate from independent information
about the galaxies. The former uses the continuum, measured through
broad band photometry over a wide wavelength range \citep{GAMA:Pan},
whereas this study targets a reduced set of age- and
metallicity-sensitive line strengths. The results are compatible, with
age differences involving pairs in lower and higher mass primary
galaxies as large as $\sim$1--2\,Gyr, although the spectral analysis is
more sensitive to small age differences. This 
environment-related effect decreases with increasing secondary mass.
The variations with respect to metallicity are negligible, although
we note that the index plots (Fig.~\ref{fig:EWs}) give a significant
separation of metallicity-sensitive indices, such as [MgFe]$^\prime$,
with respect to primary mass.  The likelihood analysis --
marginalising over all possible values of the population parameters --
washes out this information, but at the low-mass end, the equivalent
width trends (Fig.~\ref{fig:EWs}) suggest a trend towards more metal
rich satellites when located close to more massive galaxies.

Fig.~\ref{fig:Sersic} compares the structural parameters of satellite
galaxies as a function of stellar mass. We use the S\'ersic
decomposition of the surface brightness profiles presented
in \citet{Kelvin:12}. On the left, the top panel shows the projected
effective radius in physical units, and the bottom panel gives the
S\'ersic index. The sample of satellite galaxies used in the spectral
stacking -- i.e. avoiding those that featured low S/N, fringed spectra
or with n$_{\rm Q}<$3 -- is shown as grey dots. The blue and red
symbols give the average and (RMS) scatter of subsamples binned with
respect to stellar mass, corresponding to satellites close to the most
(red) or least (blue) massive primary. The aim of this figure is to
assess whether the stacked data represent structurally different
systems. The figure discards this hypothesis. Within the scatter, both
subsamples correspond to similar types of galaxies, and no significant
systematic would be expected from this issue. On the right, the histograms
show the distribution of the S\'ersic index (top) and the effective
radius (bottom) for the sample of primary (i.e. massive) galaxies.
Different histograms correspond to subsamples split with respect to
halo mass or the presence of a satellite. No significant
differences can be found, except for a weak trend towards
lower S\'ersic indices in primary galaxies without a satellite, or in
groups with lower halo masses.  The histograms show that the effects
on the stellar populations cannot be caused by a morphology-related
selection bias.
\bigskip

\subsection{Population differences in massive primary galaxies.}
\label{Ssec:PopMassive}

\begin{figure*}
\begin{center}
\includegraphics[width=14cm]{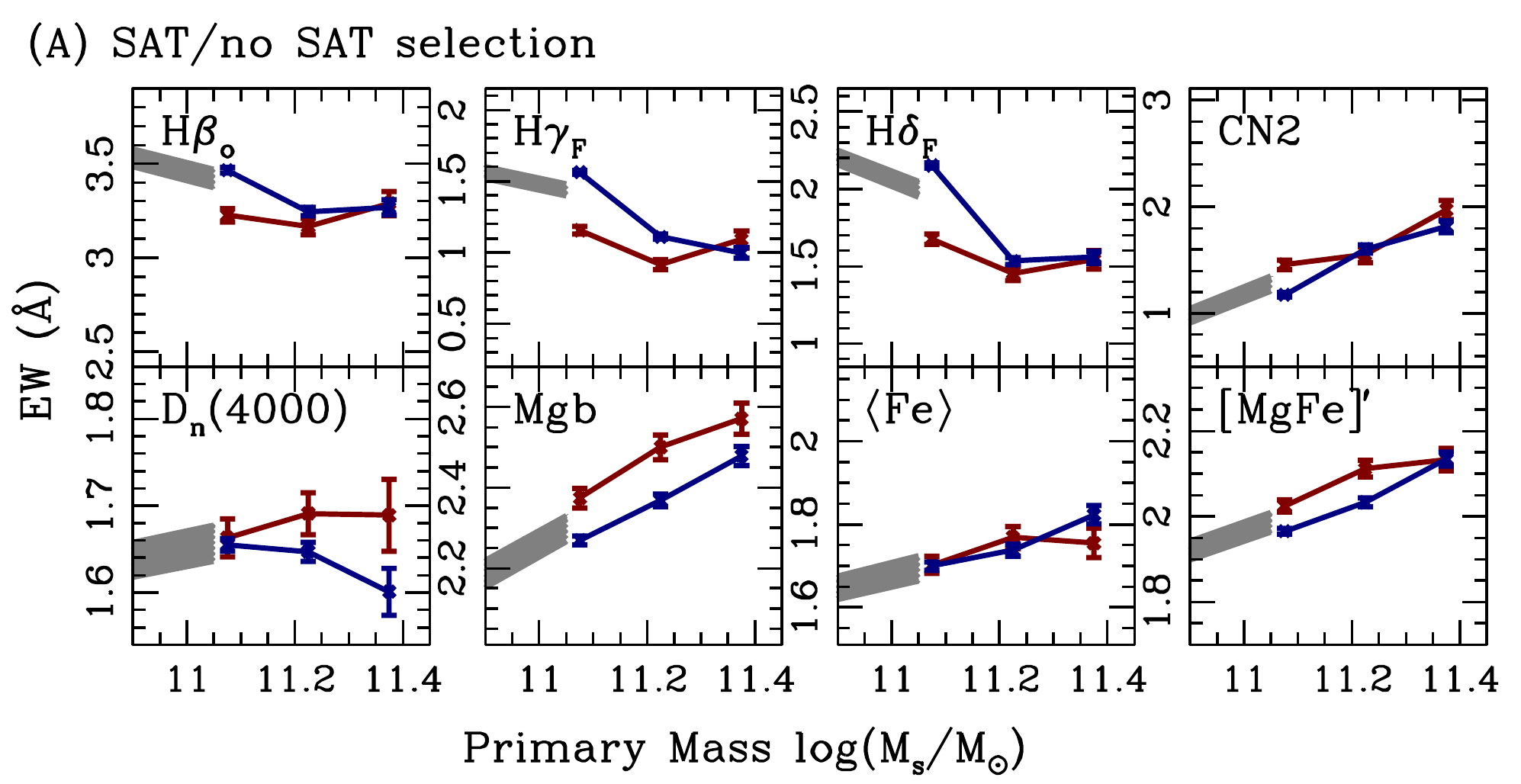}
\includegraphics[width=14cm]{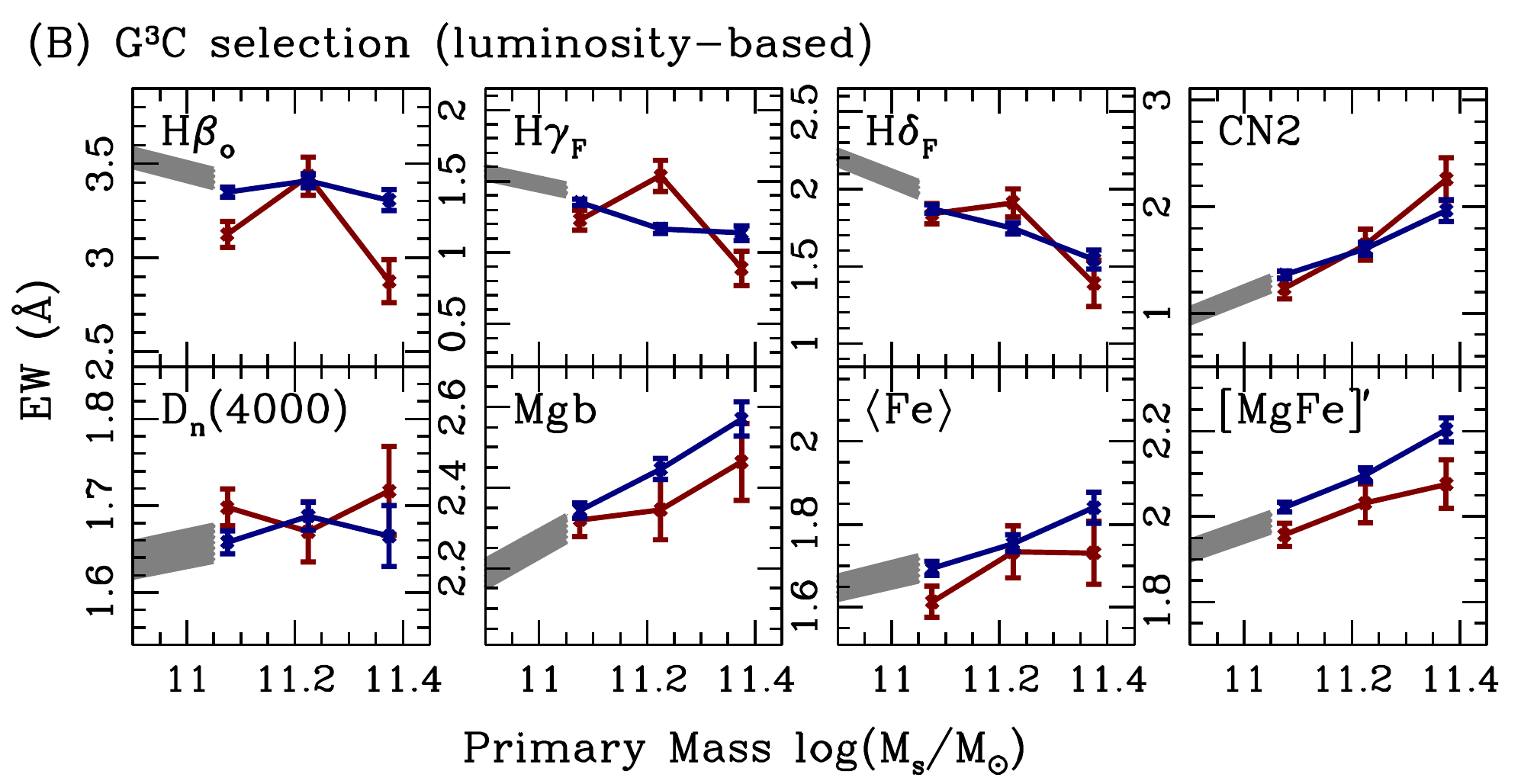}
\includegraphics[width=14cm]{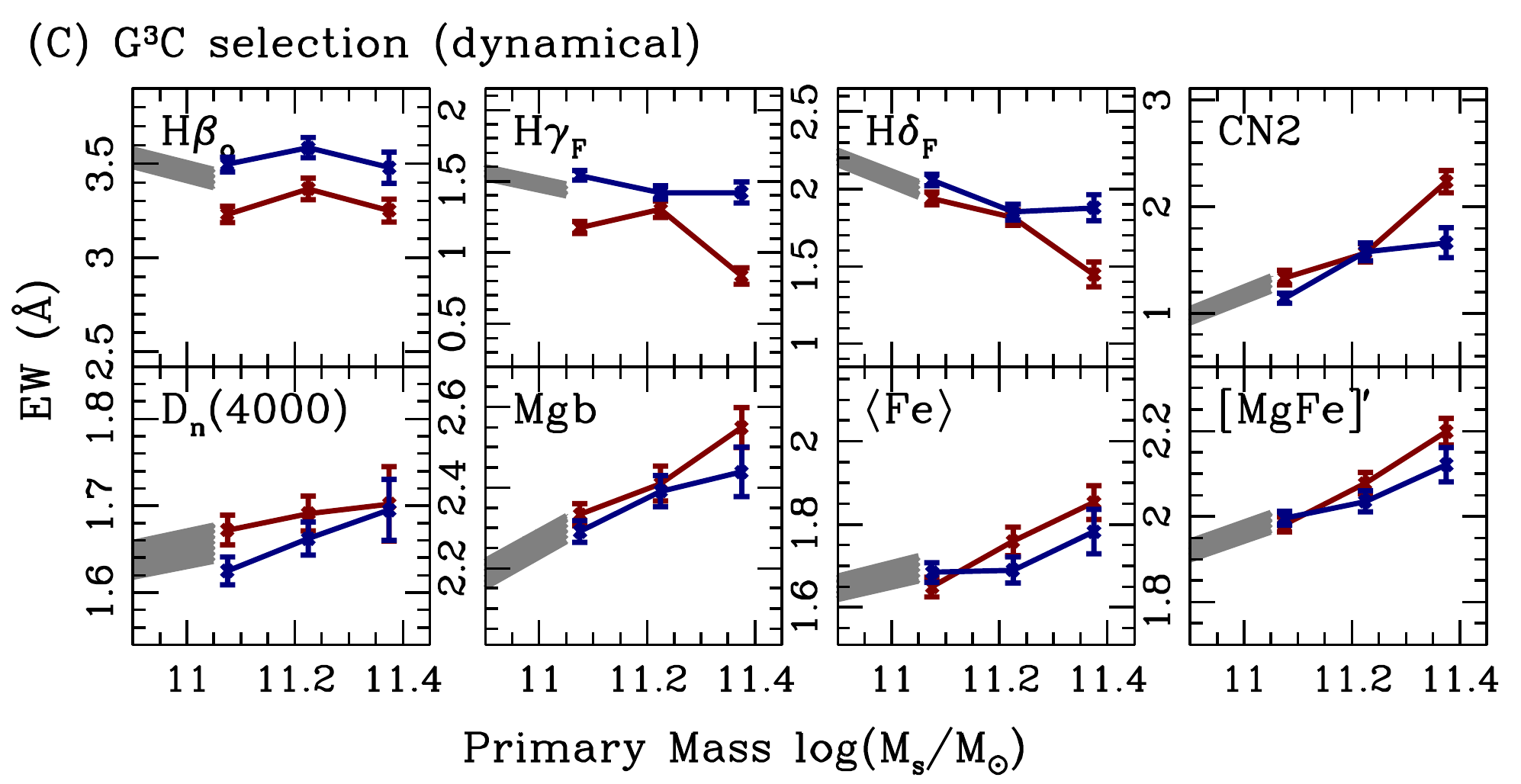}
\end{center}
\caption{Equivalent widths of stacked spectra of primary galaxies, shown as a
function of stellar mass. In (A, top) the stacking criterion is based on
whether the primary has a satellite (red) or not (blue). In (B-C, middle and bottom) the
spectra are stacked according to group mass, derived from the G$^3$C
catalogue \citep{G3C}, using either thescaling relation between halo mass
and total luminosity (B), or the dynamical mass (C).  The red (blue) lines
represent primary galaxies located in the the highest (lowest) groups.
The grey shaded region corresponds to the general trend in satellite
galaxies, as shown in Fig.~\ref{fig:EWs}. See text for details.}
\label{fig:CenEWs}
\end{figure*}

In addition to the analysis of the populations in satellites, we
extend the study to their massive companions. As these are massive
systems -- $\log$M$_s>10^{11}$M$_\odot$ -- we expect a rather
homogeneous population of red-sequence galaxies. Fig.~\ref{fig:CenEWs}
shows the trends of the EWs in stacked spectra of primary galaxies, as
a function of their stellar mass. For reference, we include the shaded
region of Fig.~\ref{fig:EWs} that represents the massive end of the
satellite sample. Analogously to the satellite sample, three different
stacking criteria are pursued: On the top panels (labelled A), the red
(blue) lines correspond to stacks of primary spectra with (without) a
nearby satellite. On the middle and bottom panels (labelled B and C),
the red (blue) lines are the results for primary galaxies in the most
(least) massive groups, according to the G$^3$C catalogue, following
the same criteria as in Fig.~\ref{fig:grid}.  The preparation of the
stacked spectra follows the same methodology as for the satellite
sample (\S\S\ref{SSec:Stack}), although the adopted fiducial velocity
dispersion is 250\,km\,s$^{-1}$. In (A) the Balmer line indices show a
trend towards older populations in galaxies with a satellite (red),
consistent with the higher 4000\AA\ break. The trend appears stronger
in D$_n$(4000).  One could expect this result from the fact that
Balmer indices are mostly sensitive to recent episodes of formation
(within $\simlt$1-2\,Gyr), whereas D$_n$(4000) varies over a wider
range of stellar ages \citep[see, e.g.,][]{Kauff:03}. However, the
interpretation of the 4000\AA\ break index gets more complicated in
old populations such as those expected in massive galaxies, where the
degeneracy with respect to metallicity is more pronounced. The
metallicity indices Mgb and [MgFe]$^\prime$ also show an increased
strength in those primary galaxies with a nearby satellite, whereas
$\langle$Fe$\rangle$ does not show any variation. In (B and C),
segregated with respect to group mass, some differences are apparent,
although weaker, except for CN2, which has a pronounced variation with
the luminosity-weighted halo masses. This result is consistent with
the environment-related trends found in this index
in \citet{Carretero:07}.  The metallicity differences (through Mgb and
[MgFe]$^\prime$) are also weak, with differences between the
methodology used to derive halo masses.
Thus, we conclude from this result that the populations of the
massive primary galaxies appear homogeneous, with a small
difference between primaries with and without satellites.
A comparison of the Mgb and the $\langle$Fe$\rangle$ line strengths
in the topmost panels of Fig.~\ref{fig:CenEWs} may suggest
an overall [Mg/Fe] enhancement in massive galaxies with a satellite.
Following the simple proxy of [Mg/Fe] presented in
\citet{FLB:13}, namely comparing the total
metallicity when constraining the populations using
either Mgb or $\langle$Fe$\rangle$ as a metallicity
indicator (instead of [MgFe]$^\prime$), gives an overall
enhanced population of the primary galaxies around
[Mg/Fe]$\sim +0.2 {\rm \ to} +0.3\, (\pm 0.1)$\,dex, but no discernible
difference -- within error bars -- between the stacks of primary
galaxies with and without satellites. We expect the
observed variations in Mgb and $\langle$Fe$\rangle$, shown in Fig.~\ref{fig:CenEWs}, to be
caused by a complex combination of age, metallicity and [Mg/Fe]
differences. The weaker trends in the group-selected stacks may be
representative, but this issue is beyond the scope of this paper,
leaving it to future work.
\bigskip

\subsection{Comparison with the close pair selection of Davies et al.}
\label{Ssec:CompDavies}

Our results provide an independent approach to the evolution of close
galaxy pairs with respect to \citet{Davies:Pairs}, who focused on star
formation diagnostics, rather than on the underlying stellar
populations. They found star formation to be suppressed in secondary
galaxies involving minor mergers. This is equivalent to the older
populations found in satellites around the most massive primary
galaxies (red lines at the low-mass end in Figs.~\ref{fig:EWs} and
\ref{fig:Age}). However, we emphasize that stellar populations provide
a cumulative (integral) picture of the past star formation history,
whereas a star formation diagnostic gives the ``instantaneous''
(differential) version of the same process, more specifically within
the last $\sim$100\,Myr.  Therefore, the study of \citet{Davies:Pairs}
is more sensitive to recent events triggering star formation, whereas
this work reveals a deeper connection with environment over longer
timescales ($\simgt 1$\,Gyr). These results are also consistent with the
radial gradients found in massive early-type galaxies, with a dominant
old and metal-poor component in the outer regions \citep[see,
e.g.,][]{FLB:11,FLB:12,Greene:15}.

\section{Summary}
\label{Sec:Summ}

Taking advantage of the uniform spatial completeness of the GAMA
survey, we select a sample of dynamically close pairs involving at
least one massive (stellar mass $\simgt 10^{11}$M$_\odot$) galaxy,
over the redshift range $0.1\leq z\leq 0.3$.  The study focuses on the
stellar populations of the satellite galaxies through a targeted set
of spectral features. Since the populations in the primary and
secondary galaxies will eventually merge, this study provides insight
on the dominant growth channel of massive galaxies during the
so-called ``second'', ex-situ phase. Due to the S/N and flux
calibration characteristics of the GAMA/AAT spectra, the analysis is
based on a continuum-subtracted version of the data, stacking
individual spectra according to two parameters: a ``local'' observable -- the stellar
mass of the satellite, and an environment-related observable, using
either the mass of the primary galaxy, or the group (i.e. dark matter halo)
mass. In addition to the well-known local trend between
age/metallicity and galaxy mass, we find a significant environmental
trend in the stellar populations, so that {\sl at fixed mass},
satellite galaxies linked to more massive primary galaxies appear older. This
trend is especially apparent at the low mass end of our sample (satellite
mass $\sim 10^{10}$M$_\odot$), where the SSP-equivalent age differences are
$\sim$1--2\,Gyr. This age difference decreases towards the more massive
satellites, and it is most significant when
considering either the mass of the primary or the group mass as the
environment-related parameter. The data cannot disentangle the effect
between these two. We emphasize that, in contrast to the recent study
of close pairs using star formation diagnostics \citep{Davies:Pairs},
this work focuses on the stellar populations, providing insight into
the star formation processes over longer timescales. The consistency
of these results reinforce the idea that galaxy-related processes due
to the primary must play an important role on the observed
differences, and that the trends with respect to group halo mass may be inherited
from the intrinsic correlation between the two.  Therefore, our
results are consistent with the general picture of galactic
conformity \citep{Weinmann:06}. The stellar mass of a galaxy
is robustly found as the main indicator of the properties of its underlying
stellar populations. However, at fixed stellar mass, the population of
satellite galaxies have more in common with the corresponding central
in its group. In this work, this conformity appears in the age of the
stellar populations: if the primary galaxy is older (roughly more
massive), the age of the secondary is older than that of another
secondary {\sl with the same stellar mass}, orbiting a less massive,
thus younger, primary. Such a result is also consistent with the observed
lack of age gradients in giant early-type galaxies \citep{FLB:11,FLB:12}.
These observational trends should provide useful constraints for
numerical simulations of galaxy formation, where the internal age and
metallicity gradients are very sensitive to the sub-grid
physics \citep[see, e.g.][]{Hirschmann:15}.

\section*{Acknowledgements}
IF gratefully acknowledges support from the AAO through their
distinguished visitor programme in 2014 and 2016, as well as support
from ICRAR and the Royal Society.  MLPG acknowledges CONICYT-Chile
grant FONDECYT 3160492.MSO acknowledges the funding support form the
Australian Research Council through a Future Fellowship
(FT140100255). Funding for SDSS-III has been provided by the Alfred
P. Sloan Foundation, the Participating Institutions, the National
Science Foundation, and the U.S. Department of Energy Office of
Science. The SDSS-III web site is http://www.sdss3.org/. GAMA is a
joint European-Australasian project based around a spectroscopic
campaign using the AAT. The GAMA input catalogue is based on data
taken from the SDSS and the UKIRT Infrared Deep Sky
Survey. Complementary imaging of the GAMA regions is being obtained by
a number of independent survey programmes including GALEX MIS, VST
KIDS, VISTA VIKING, WISE, Herschel-ATLAS, GMRT and ASKAP providing UV
to radio coverage. GAMA is funded by the STFC (UK), the ARC
(Australia), the AAO and the participating institutions. The GAMA web
site is http://www.gama-survey.org/.


\label{lastpage}

\end{document}